\documentclass[12pt]{article}

\usepackage{natbib,hyperref}

\usepackage{graphicx,bm,colonequals,amsmath,amssymb,url,xcolor,bbm, mathtools}
\usepackage{array,tabularx,multirow}
\usepackage{caption,subcaption}
\usepackage[utf8]{inputenc}
\usepackage{enumitem}
\usepackage{soul} 
\usepackage[normalem]{ulem}

\usepackage{multirow}

\newcommand{\blind}{1}

\newcommand{\bh}{\mathbf{h}}
\newcommand{\bb}{\mathbf{b}}
\newcommand{\bs}{\mathbf{s}}
\newcommand{\bI}{\mathbf{I}}

\newcommand{\bfzero}{\mathbf{0}}
\newcommand{\bftheta}{\bm{\theta}}
\newcommand{\bfeta}{\bm{\eta}}
\newcommand{\bfSigma}{\bm{\Sigma}}
\newcommand{\bfLambda}{\bm{\Lambda}}

\newcommand{\GP}{\mathcal{GP}}
\newcommand{\normal}{\mathcal{N}}

\newcommand{\domain}{\mathcal{D}}

\newcommand{\mra}{{$M$-RA }}

\begin{document}
	
	\def\spacingset#1{\renewcommand{\baselinestretch}%
		{#1}\small\normalsize} \spacingset{1}
	
	
	\if1\blind
	{
		\thispagestyle{empty} \baselineskip=28pt \vskip 5mm
		\begin{center} {\LARGE{\bf Nonstationary Spatial Modeling of Massive Global Satellite Data}}
		\end{center}
		
		\begin{center}\large
			Huang Huang\footnote[1]{
				\baselineskip=11pt
				Statistics Program,
				King Abdullah University of Science and Technology, Thuwal, 23955-6900, Saudi Arabia. huang.huang@kaust.edu.sa
			}, 
			Lewis R. Blake\footnote[2]{
				\baselineskip=11pt 
				Department of Applied Mathematics and Statistics,
				Colorado School of Mines, Golden, CO, 80401, USA. lblake@mines.edu, hammerling@mines.edu
			}, 
			Matthias Katzfuss\footnote[3]{
				\baselineskip=11pt 
				Department of Statistics,
				Texas A\&M University, College Station, TX, 77843, USA. katzfuss@tamu.edu
			}, and Dorit M. Hammerling\textsuperscript{2}
		\end{center} 
	} \fi
	
	
	\if0\blind
	{
		\thispagestyle{empty} \baselineskip=28pt \vskip 5mm
		\begin{center}
			{\LARGE\bf Nonstationary spatial modeling of massive global satellite data}
		\end{center}
		\medskip
	} \fi
	
	\bigskip
	\begin{abstract}
		Earth-observing satellite instruments obtain a massive number of observations every day. For example, tens of millions of sea surface temperature (SST) observations on a global scale are collected daily by the Moderate Resolution Imaging Spectroradiometer (MODIS) instrument. Despite their size, such datasets are incomplete and noisy, necessitating spatial statistical inference to obtain complete, high-resolution fields with quantified uncertainties. Such inference is challenging due to the high computational cost, the nonstationary behavior of environmental processes on a global scale, and land barriers affecting the dependence of SST.
		In this work, we develop a multi-resolution approximation (\mra\!\!) of a Gaussian process (GP) whose nonstationary, global covariance function is obtained using local fits. 
		The \mra requires domain partitioning, which can be set up application-specifically. In the SST case, we partition the domain purposefully to account for and weaken dependence across land barriers.
		Our \mra implementation is tailored to distributed-memory computation in high-performance-computing environments.
		We analyze a MODIS SST dataset consisting of more than 43 million observations, to our knowledge the largest dataset ever analyzed using a probabilistic GP model. We show that our nonstationary model based on local fits provides substantially improved predictive performance relative to a stationary approach. 
	\end{abstract}
	
	\noindent%
	\vfill
	
	\newpage
	\spacingset{1.45} 

\section{Introduction\label{sec:intro}}

NASA's large fleet of earth-observing satellites obtain a massive number of observations each day. After pre-processing and retrieval, this results in billions of geophysically calibrated and georeferenced swath measurements on a global scale, referred to as Level-2 data by NASA. Measurements of a particular variable are typically noisy and incomplete, leaving big gaps in daily Level-2 maps, which prohibit the optimal use of these important data sources and, in turn, prevent many scientific problems from being resolved.
Hence, the value and applicability of these data to answer important science questions can be greatly improved if the true, complete spatial fields are inferred optimally at high resolution, and uncertainties are rigorously quantified.

Here we focus on sea-surface temperature (SST). As the oceans cover more than 70\% of Earth's surface and are critical for Earth's life and ecosystems, SST is essential for hurricane prediction, weather forecasting, and climate-change research \citep{o2019observational}.
The Moderate Resolution Imaging Spectroradiometer (MODIS) on the NASA Aqua satellite provides tens of millions of Level-2 SST observations per day at a very high $1\textrm{km}^2$ spatial resolution on a global scale. These observations exemplify the statistical challenges associated with analyzing Level-2 satellite data: The datasets are massive, global, and the underlying fields nonstationary. In addition, land masses provide barriers, which can reduce spatial dependence at given distances and thus provide further sources of nonstationarity.

Current practices for turning Level-2 satellite data into complete, gridded data products, referred to as Level-3 data, are often relatively simple due to convenience, computational limitations, or the need for very general products that are useful to a broad user base. In many cases, including NASA's official Level-3 data products, data are gridded by simply computing averages over coarse spatial and temporal grids. This approach is fast and intuitive but also leads to a loss of fine-scale features and opaque uncertainties for further use of the data, making it difficult or impossible to answer some scientific questions.

Spatial statistical approaches provide a formal toolbox for analyzing spatial data, enabling more accurate spatial predictions and rigorous assessment of uncertainties. 
Based on a statistical model parameterizing the spatial dependence structure, estimation of parameters and spatial prediction (``kriging'') can be carried out \citep[e.g.,][]{Cressie1993}. So-called spatial and spatio-temporal hierarchical models \citep[e.g.,][]{Cressie2011} are essentially flexible extensions of kriging that can handle a variety of different types of observations. However, most of these models are computationally infeasible for very large satellite datasets.
For oceanic data analyses, optimal interpolation \citep{Bretherton1976} is still the most common algorithm for blending, say SST and data obtained by different instruments \citep{Chao2009}. 
Optimal interpolation is essentially a special case of kriging, but it has several drawbacks in its typical form. For example, it is computationally infeasible for large datasets and lacks suitable models for spatial nonstationarity, the latter of which is important for many environmental variables, including SST, known to be smoother and less variable offshore than nearshore \citep[e.g.,][]{Chao2009}.

Recent endeavors to scale spatial statistical approaches to large spatial data were reviewed by \citet{Heaton2017}, which included a numerical comparison on a satellite dataset with about 0.15 million observations.
\citet{SFCB2019} proposed a spatial process model as the sum of a low-rank Gaussian predictive process~\citep{Banerjee2008} and a nearest-neighbor Gaussian process~\citep{DBFG2016} approximation of a stationary covariance using a shared-memory parallelization scheme (OpenMP); their approach was applied to remote-sensing data sets of around 17 million observations. 
\citet{FDB2020} implemented an \texttt{R}~\citep{R2020} package for nearest-neighbor Gaussian process approximations of stationary covariances with OpenMP and showed an application example of around 38 million observations. 
\cite{ZR2020} considered nonstationarity modeling via a multi-scale random process. They projected the process to a triangulation grid and induced conditional independence so that the precision is sparse and inference is feasible for large datasets. In the application, they sampled one million observations from global SST satellite data and performed the inference using shared-memory parallelism.
\citet{AP2020} proposed the nonstationary \mra framework in the spatio-temporal setting and applied their \texttt{R} implementation to moderately large spatio-temporal SST datasets with around 0.6 million observations.

Of the many SST products based on satellite data provided by the Group for High Resolution Sea Surface Temperature (GHRSST), the multi-scale ultra-high-resolution (MUR) SST \citep{Chin2013} is possibly the most advanced product currently available. It employs multi-resolution variational analysis (MRVA), which uses wavelet-based multi-scale signal expansion to address the irregularity in measurement locations and scale-dependent issues \citep{Chin1998}. However, the MRVA uses basis functions of a fixed shape, which for SST can result in over-smoothing in coastal areas and under-smoothing off-shore. Also, additional approximations are needed to accommodate the irregularly spaced data and the uncertainty estimates, which are only point-wise.

Here, we provide an efficient method for analyzing global Level-2 satellite data based on the multi-resolution approximation ($M$-RA) for spatial processes first introduced in \citet{Katzfuss2015} and \citet{Katzfuss2017b}, which uses a large number of basis functions to capture spatial variation at all scales while being highly computationally efficient. In contrast to related multi-resolution wavelet models (e.g., the MRVA), the $M$-RA is directly applicable to irregularly spaced observations, allows proper probabilistic inference, and the basis functions can adjust flexibly to spatially-varying dependence structure.

In this work, we extend this basic \mra approach in several ways. We combine the \mra with a nonstationary, global covariance function; specifically, we employ a nonstationary Mat\'ern covariance in $\mathbb{R}^3$ with spatially varying parameters \citep{Paciorek2006} on the 2-sphere embedded in $\mathbb{R}^3$, with parameters varying according to local fits of isotropic Mat\'ern covariances. 
We also develop an ocean-specific scheme, where we partition boundaries and place the basis-function  for the \mra in such a way that we can specifically account for, and weaken, dependence across land barriers emulating the dependence structure of SST across different ocean basins.
On the computational side, we provide and describe an implementation of the \mra tailored to distributed-memory computation in high-performance-computing environments that can split the computing task efficiently between many computational nodes; further details of this implementation can be found in several technical notes \citep{HBH2019,BHVH2019, BHVH2021}.

The resulting method is well suited to the analysis of Level-2 satellite data, enabling the nonstationary analysis of massive, global data, while fully quantifying (even joint) uncertainties and capturing dependence at all spatial scales.
Our method is applicable to a broad array of Level-2 satellite data across all NASA Earth science focus areas, where it will allow scientific questions to be answered more accurately and precisely.

In this work, we focus on one important application: providing a new high-resolution SST product based on Level-2 MODIS data with more than 43 million observations. On the Cheyenne supercomputer at the National Center for Atmospheric Research, we obtain the full posterior predictive distribution of SST on a fine grid in a computationally highly efficient manner, as well as with high statistical efficiency. Our flexible nonstationary model captures the spatially varying dependence structure, which leads to substantially improved prediction relative to the stationary model~\citep{HBH2019, BHVH2019, BHVH2021} in terms of both accuracy and uncertainty quantification. Thus, our approach is highly useful for oceanographic research that requires fine-scale joint distributions. For example, the magnitude of gradients in ultra-high-resolution SST products is strongly related to the intensity of the upwelling of cold, nutrient-rich water from depth \citep{Vazquez2017}, and thus accurate gradient calculations allow proper investigation of changes in upwelling and the impact on coastal fisheries due to climatic changes.
To our knowledge, our MODIS SST dataset is the largest  ever analyzed using a probabilistic Gaussian process model, especially with a nonstationary covariance structure.

The remainder of the manuscript is organized as follows. Section~\ref{sec:model} introduces the nonstationary \mra model and our distributed-computing implementation.
Section~\ref{sec:data} discusses the SST data, important aspects related to  performing predictions, the nonstationary prediction results with the improvement over a stationary model shown by several assessment metrics, and the complete and high-resolution SST product. Section~\ref{sec:discuss} concludes and points out future research directions.




\section{Modeling\label{sec:model}}
    \subsection{A Brief Review of the \mra Model}
\begin{figure}
	\centering
	\includegraphics[width =.8\linewidth]{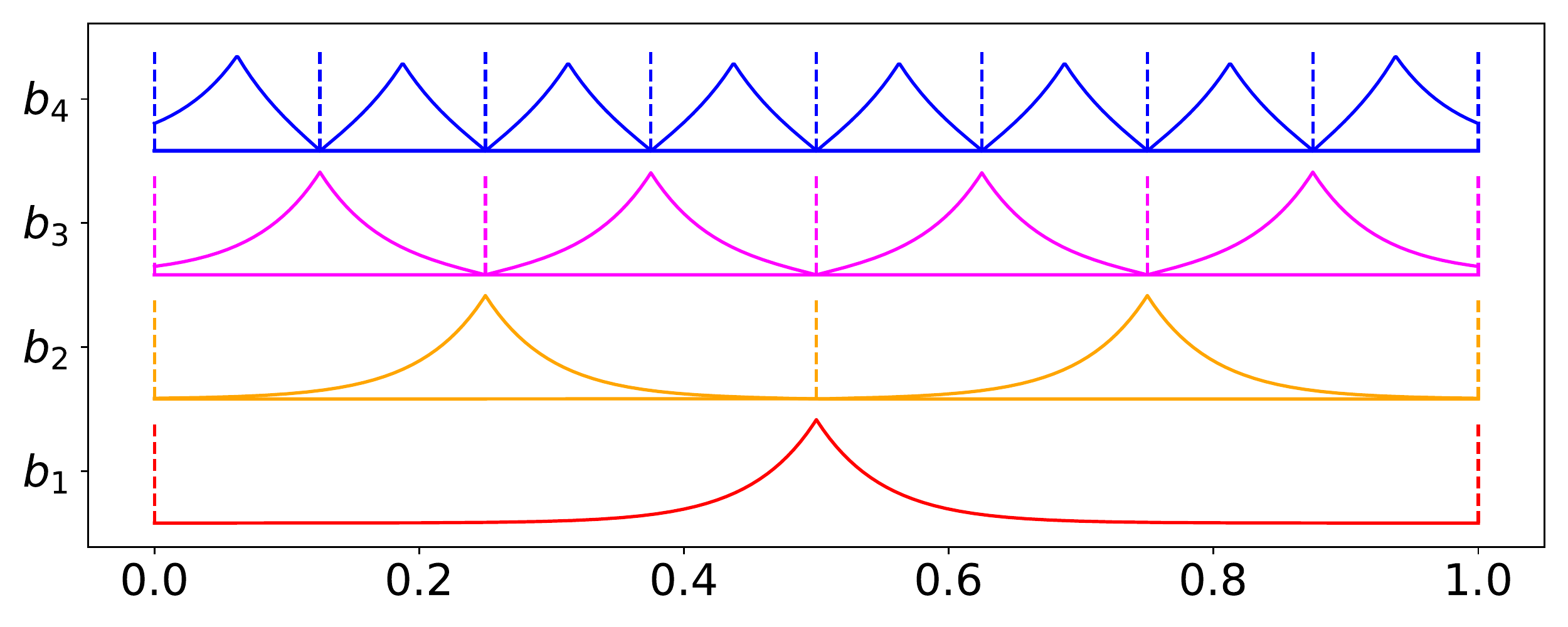}
	\caption{A toy example of basis functions $\bb_m(\cdot)$ in the \mra on a one-dimensional spatial domain $\domain = [0,1]$ with $M=4$ levels, $J=2$ subregions per region (vertical lines), and $r=1$ knot per region.}
	\label{fig:toymradata}
\end{figure}

    In this section, we provide a brief review of the \mra method.
For more details on the \mra model, the reader is directed to \cite{Katzfuss2015}. Assume that $y(\cdot) \sim \GP\big(0,C(\cdot,\cdot)\big)$ is a Gaussian process (GP) with covariance function $C(\cdot,\cdot)$. Direct inference based on $n$ observations of this GP is computationally infeasible for large $n$. Similar to wavelet approaches, the $M$-RA process is specified as a linear combination of basis functions at multiple levels of spatial resolution, which can capture spatial structure from very fine to very large scales. The $M$-RA requires a hierarchical partitioning of the domain $\domain$, in which we recursively split each region into $J$ (e.g., $J=2$) subregions, up to some level $M$ (e.g., $M=10$). Within each region, we specify a grid consisting of a small number $r$ (e.g., $r=30$) of locations (called knots). These knots will form the centers of basis functions, whose shape is then determined iteratively by the predictive process \citep{Banerjee2008}. The basis functions at each resolution are approximately optimal, in that the predictive process based on $r$ knots is the Nystr\"om approximation of the first $r$ terms in the Karhunen--Lo\`eve expansion of the process to be approximated \citep{Sang2012}. 
More precisely, the \mra process is a combination of the basis functions at all $M$ resolutions:
\begin{equation}
\label{eq:mradef}
  y_{M\textrm{-RA}}(\bs) = \sum_{m=1}^M \bb_m(\bs)'\bfeta_m, \quad \bs \in \domain,
\end{equation}
where for each resolution $m=1,\ldots,M$, $\bb_m(\cdot)$ is a vector of $p_mr$ spatial basis functions, $\bfeta_m \stackrel{ind.}{\sim}\normal_{p_mr}(\bfzero,\bfLambda_m)$ is a vector of corresponding basis-function weights, and $p_m$ is the number of regions at resolution $m$.
Figure \ref{fig:toymradata} illustrates the \mra in a simple toy example.

One important property of the $M$-RA is that both the basis functions $\bb_m(\cdot)$ and the covariance matrix of the weights, $\bfLambda_m$, are automatically chosen to optimally approximate the true process $y(\cdot)$ and its covariance function $C(\cdot,\cdot)$. Because $C(\cdot,\cdot)$ can be any covariance function suitable for the application or data of interest, the basis-function model is easily interpretable and highly flexible (e.g., allowing the dependence structure of the process of interest to vary over the spatial domain).
Another property, crucial for computational feasibility, is that for a larger $m$, the number of basis functions in $\bb_m(\cdot)$ increases while the support of each function decreases. This is done in such a way that any location $\bs \in \domain$ is only covered by $r$ basis functions per resolution (i.e., $\bb_m(\bs)$ has only $r$ nonzero elements), and thus $Mr$ basis functions in total. Similarly, $\bfLambda_m$ is a block-diagonal matrix of increasing dimension for increasing $m$, but the blocks along the diagonal are only of size $r \times r$. This block-sparse form ensures that all computations involving the $M$-RA are highly scalable for massive datasets.

The \mra can also be viewed as a special case of an ordered conditional approximation \citep{Vecchia1988}, resulting in a block-sparse inverse Cholesky factor \citep{Katzfuss2017a,Jurek2018}.

    \subsection{Inference via Distributed Computing}
    In general, \mra inference proceeds in two stages. First, the prior covariance matrices $\bfLambda_m$ and the basis functions $\bb_m(\cdot)$ (at knot/observation locations) are computed sequentially for $m=1,2,\ldots,M$. Second, the posterior distributions of the basis-function weights $\bfeta_m$ given the data are computed sequentially in the opposite direction, for $m=M,M-1,\ldots,1$.

\begin{figure}[!t]
    \centering
    \includegraphics[width=\textwidth]{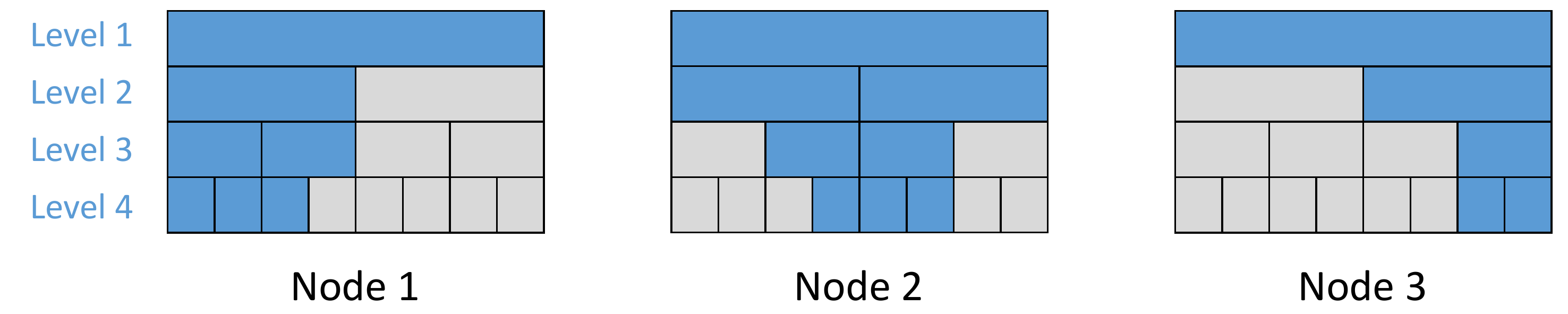}
    \caption{Working regions for each node, indicated in blue.}
    \label{fig:dc-regions}
\end{figure}

When the data size is large, even though the \mra framework hierarchically imposes low-rank representation and block-independence structure within each subregion at each resolution (which eases the computational burden and memory storage requirement), performing predictions is still intractable if only one computer is used.
Therefore, a distributed-computing implementation is a must when building a high-resolution product from massive satellite observations.
For ease of presentation, we explain the distributed computing scheme conceptually on a one-dimensional domain with the same setting as in Figure~\ref{fig:toymradata}, where $M=4$ and $J=2$.
Additional technical details are provided in a tech note \citep{HBH2019}.
We assume a cluster of three nodes is available, where each node has its own memory and computing cores, and communication among nodes is possible via networks.
In this example, we have eight regions at the finest level. We assign similar numbers of regions at the finest level to each node and subsequently all ancestral regions at coarser levels. Figure~\ref{fig:dc-regions} shows the working regions for each node.

\begin{figure}[ht!]
    \centering
    \includegraphics[width=\textwidth]{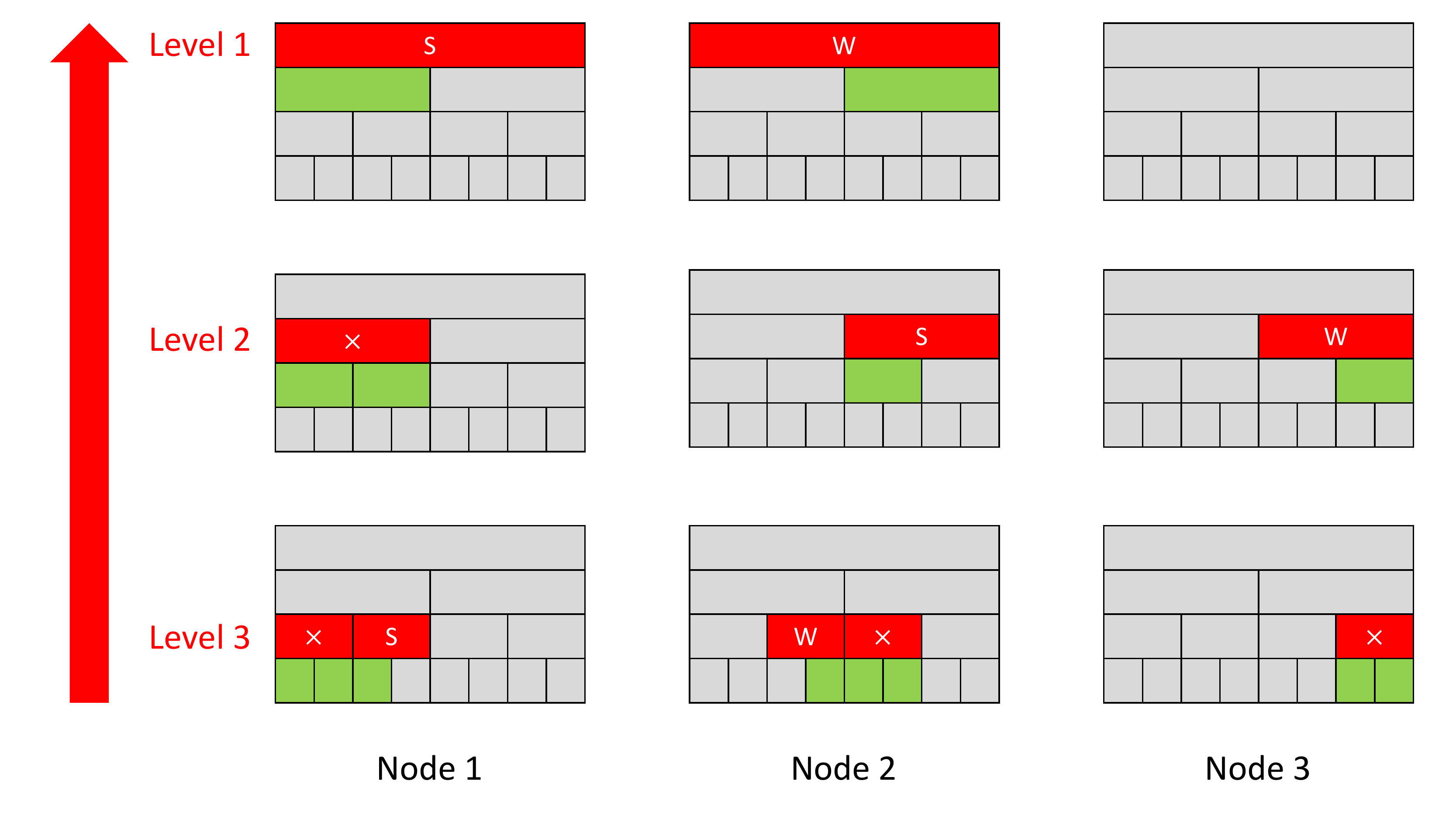}
    \caption{Summary of posterior computations.
    At each computing level, green indicates working regions at the previous level; red with a cross indicates regions where this node handled all children, so no synchronization is needed; red regions with ``S'' or ``W'' require synchronization, where ``S'' represents the “supervisor” that receives data and ``W'' is the ``worker'' that sends data.}
    \label{fig:dc-posterior}
\end{figure}

Several prior quantities related to the covariance among knots need to be computed and stored. 
These computations go from $m=1$ to 4, and at each level $m$, all computations in different regions are independent and can be executed in parallel.
However, the computations for posterior distributions given the data from $m=3$ to 1 are more complicated. 
Figure~\ref{fig:dc-posterior} summarizes different computation scenarios when obtaining the posterior.

When $m=3$, relevant quantities in the first region for Node 1 (red with a cross) can be obtained without the need to communicate with other nodes, because both its children at level 4 were handled in Node 1. The second region for Node 1 needs synchronization with Node 2, because some of its children were handled by Node 1 and some by Node 2. Node 2 sends previously computed quantities for its second child at level 4 to Node 1, so Node 1 has all the needed information to compute the posterior for the second region (red with a ``S''). After the synchronization, the second region (red with a ``W'') for Node 2 is eliminated, because Node 1 has all its information, and it is no longer green when $m=2$ for Node 2.
Similar computations are executed until we finish all the computations at level~1. 

For predictions, we then need to proceed from Level 1 to $M$, again calculating relevant quantities, and the prediction distributions for new locations are obtained at level $M$. Computations for different regions at the same level are executed in parallel and only depend on ancestral regions. 

We use Message Passing Interface (MPI) to operate the synchronization in \textsf{C}\texttt{++}.
We also use OpenMP to execute parallel computations of different regions at the same level for further computing acceleration.
    
    \subsection{Nonstationarity Modeling\label{subsec:nonstat_modeling}}
    Previous work has shown the excellent prediction performance of the \mra model with a stationary model for very large datasets~\citep{HBH2019, BHVH2019, BHVH2021}. 
However, for massive datasets on the globe, the stationarity assumption typically does not hold and flexible nonstationary models are required.
Pioneering works have investigated the extension of the \mra to account for nonstationarity. For example, \citet{BBN2021} proposed a mixture \mra model where a shrinking prior is induced as one component for the \mra basis function weights in a Bayesian framework. 

In this work, we directly extend the underlying covariance model to be nonstationary and provide high-performance computing implementation for large datasets. We build the nonstationary covariance model based on the approach proposed by \cite{Paciorek2006}. The closed form for the nonstationary correlation function $R^{NS}(\bs, \bs^\prime)$ on $\mathbb{R}^p$, $p \in \mathbbm{N}_+$ is,
\begin{equation}\label{eq:paciorek_and_schervisch}
R^{NS}(\bs, \bs^\prime) = \lvert\bfSigma_\bs \rvert^{\frac{1}{4}} \lvert\bfSigma_{\bs^\prime} \rvert^{\frac{1}{4}} \left\vert \frac{\bfSigma_\bs + \bfSigma_{\bs^\prime}}{2} \right\vert^{-\frac{1}{2}}R^{S}\left( \bs-\bs'; \dfrac{\bfSigma_\bs + \bfSigma_{\bs^\prime}}{2}\right),
\end{equation}

\begin{sloppypar}
\noindent where $\bfSigma_\bs$ is the covariance matrix of a Gaussian kernel centered at $\bs$, and $R^S(\bh;\bfSigma)$ is a stationary correlation function in Euclidean space.
We consider a stationary Mat\'ern correlation function  
$R^{S}(\bh; \bfSigma) = \big(\sqrt{2\nu\bh^\top\bfSigma^{-1}\bh}\big)^\nu K_\nu\big(\sqrt{2\nu\bh^\top\bfSigma^{-1}\bh}\big) / \{\Gamma(\nu)2^{\nu-1}\}$, where $K_\nu(\cdot)$ is the modified Bessel function of the second kind with order $\nu$.
It is also possible to allow the smoothness parameter $\nu$ to vary spatially. However, identifiability issues are noticed when $\bfSigma_\bs$ and $\nu$ are both spatially varying~\citep{AS2011}.

With this approach, we can model our data as points on the sphere using chordal distance to yield a valid nonstationary covariance on the sphere, which means $\bs \in \domain = \mathbb{R}^3$.
Applying spatially varying partial sill parameters $\sigma^2(\bs)$, nugget parameters $\tau^2(\bs)$, and range parameters $\beta(\bs)$ determining the isotropic kernel $\bfSigma_\bs = \beta^2(\bs) \bI_3$, we have all the parameters $\bftheta(\bs)=\{\sigma^2(\bs),\beta(\bs),\tau^2(\bs),\nu\}$ that need to be inferred, and the nonstationary covariance model is
\begin{equation}
\label{eq:nscov}
\begin{array}{rcl}
C(\bs,\bs';\bftheta(\bs))
&=
&
\sigma(\bs)\sigma(\bs')
\left(\dfrac{2\beta(\bs)\beta(\bs')}
{\beta^2(\bs)+\beta^2(\bs')}\right)^{3/2}
\left(
  \dfrac
    {2\sqrt{\nu}\|\bs-\bs'\|}
    {\sqrt{
        \beta^2(\bs)+\beta^2(\bs')
        }
    }
\right)^\nu\times\\
&&
K_\nu
\left(
  \dfrac
    {2\sqrt{\nu}\|\bs-\bs'\|}
    {\sqrt{
            \beta^2(\bs)+\beta^2(\bs')
        }
    }
\right)/ \{\Gamma(\nu)2^{\nu-1}\}
+
\tau^2(\bs)\mathbbm{1}_{\{\bs=\bs'\}},
\end{array}
\end{equation}
where
$\mathbbm{1}_{\{\}}$ denotes the indicator function (see Appendix for derivation).
\end{sloppypar}
To be able to calculate the nonstationary covariance between any pair of locations shown in Equation~\eqref{eq:paciorek_and_schervisch}, estimates of $\bftheta(\bs)$ are required for any location $\bs\in\domain$. 
We propose to compute local estimates on a regular grid; these estimates can then be smoothed by regressing them on a set of basis functions to obtain a smooth map $\bftheta(\bs)$. 
More specifically,
after specifying a grid over the study area, a local estimate of $\bftheta(\bs^\ast)$ is obtained at each grid point $\bs^\ast$ by the following procedure (which can be carried out in parallel):
\begin{enumerate}
    \item We build two boxes centered at $\bs^\ast$ with different sizes, $B_1(\bs^\ast)$ and $B_2(\bs^\ast)$, where $B_2(\bs^\ast)$ is larger than $B_1(\bs^\ast)$.
    \item We randomly choose up to $N_s$ of the observations in $B_1(\bs^\ast)$ to obtain information on the local short-range dependence, and $N_b$ observations in $B_2(\bs^\ast)\backslash B_1(\bs^\ast)$ that contain information on the long-range dependence. 
    \item We find the maximum likelihood estimates of the partial sill $\sigma^2$, range $\beta$, smoothness $\nu$, and nugget $\tau^2$ in the stationary Mat\'ern covariance function  $\mathcal{M}(\bs-\bs^\prime) = \sigma^2\big(\sqrt{2\nu}\|\bs-\bs^\prime\|/\beta\big)^\nu K_\nu\big(\sqrt{2\nu}\|\bs-\bs^\prime\|/\beta\big) / \{\Gamma(\nu)2^{\nu-1}\}+\tau^2\mathbbm{1}_{\{\bs=\bs^\prime\}}$ based on the $N_s+N_b$ chosen observations. We then assign the obtained parameter estimates to the lattice point $\bs^\ast$.
\end{enumerate}

Once estimates are obtained for each grid point, we smooth the local parameter estimates and extend them to locations not on the grid by regressing the estimates onto a set of Wendland basis functions.
We place $N_w$ compactly supported Wendland basis functions centered at $\tilde\bs$ with the length of support $\ell$ as follows,
\[
\textrm{WL}(\bs;\tilde\bs) = 
\frac{1}{3}
\bigg(1 - \frac{\|\bs-\tilde\bs\|}{\ell}\bigg)^6
\bigg(
    35\Big(\frac{\|\bs-\tilde\bs\|}{\ell}\Big)^2+
    18\frac{\|\bs-\tilde\bs\|}{\ell}+
    3
\bigg)\times\mathbbm{1}_{\{\|\bs-\tilde\bs\|<\ell\}},
\]
We place $\tilde\bs$ in the study area according to an icosahedral strategy, which ensures equivalent spacing among $\tilde\bs$ across the domain. 
For each parameter estimation, we can add a Lasso penalty to avoid overfitting, and the value of $N_w$, the penalty parameter for the Lasso regularization, and the support length $\ell$ can be optimized via cross-validation with respect to the mean squared prediction error (MSPE). 

\section{MODIS Sea Surface Temperature Analysis\label{sec:data}}
    \subsection{Data Description}
    We use GHRSST Level-2P Global Sea Surface Skin Temperature data from the MODIS instrument on the NASA Aqua satellite (GDS2), which are available from NASA's Physical Oceanography Distributed Active Archive Center (PODAAC) at \url{https://podaac.jpl.nasa.gov/dataset/MODIS_A-JPL-L2P-v2019.0}.  Both daytime and nighttime measurements collected at a $1\textrm{km}^2$ resolution are provided. For exploratory purposes, we analyze the daytime and nighttime data of days 1 through 7 of January, April, October, and December to inspect seasonal SST fluctuations and determine when the highest number of observations are available. We notice that the data in these dates tend to have more observations in the nighttime than daytime. Examples of the nighttime SST are shown in Figure \ref{fig:globalL2Daily2019_low_res} in the Supplementary Material. 
The raw data contain a quality level flag from 0 to 5, indicating the accuracy of the recorded observations. The User Guide in the data product suggests that observations with quality levels 2 or greater are usable, so we eliminate all the observations with  quality levels less than 2.

\begin{figure}[!t]
    \centering
    \includegraphics[width=\textwidth]{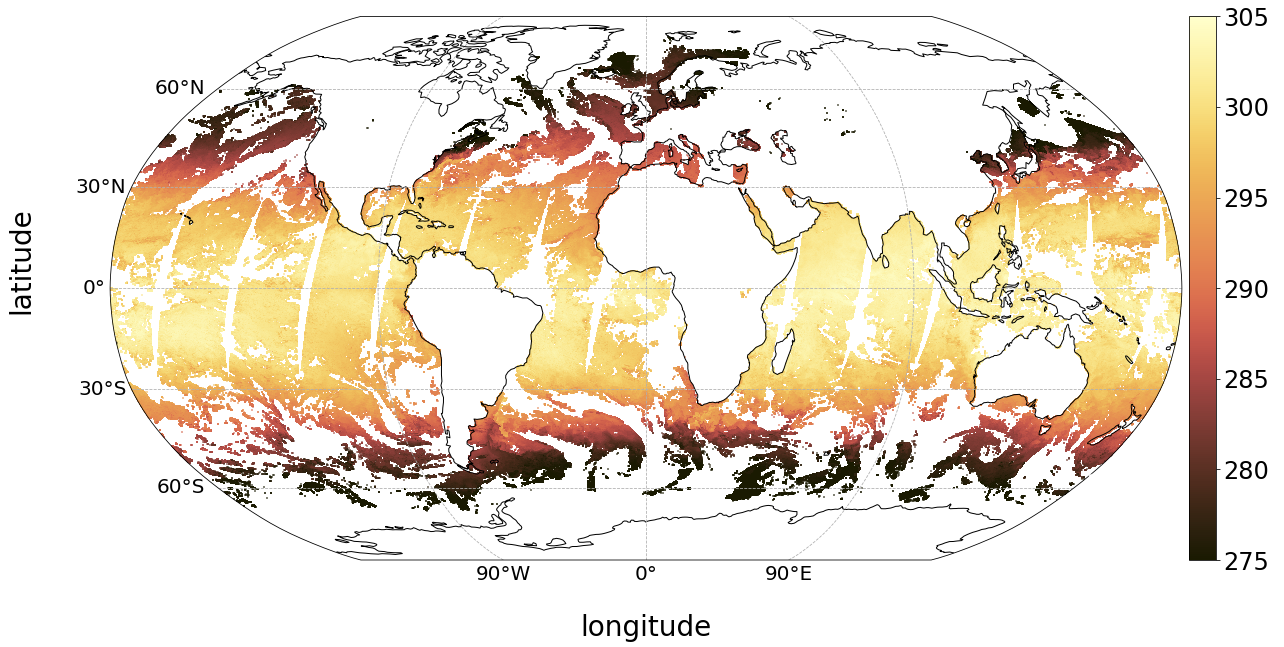}
    \caption{Global nighttime SST on April 7, 2019, with $n = 45{,}057{,}134$ observations.}
    \label{fig:dataset}
\end{figure}

We select the date with the largest amount of available data, which is nighttime, April 7, 2019, as the primary data set of consideration for the remainder of our analysis, illustrated in Figure~\ref{fig:dataset}.
Nighttime SST from April 1 to 6, 2019, will be used as additional reference data sources for the covariance modeling.



    \subsection{Removing Mean Structure\label{subsec:mean}}
\begin{figure}[!htb]
    \centering
    \includegraphics[width=0.75\textwidth]{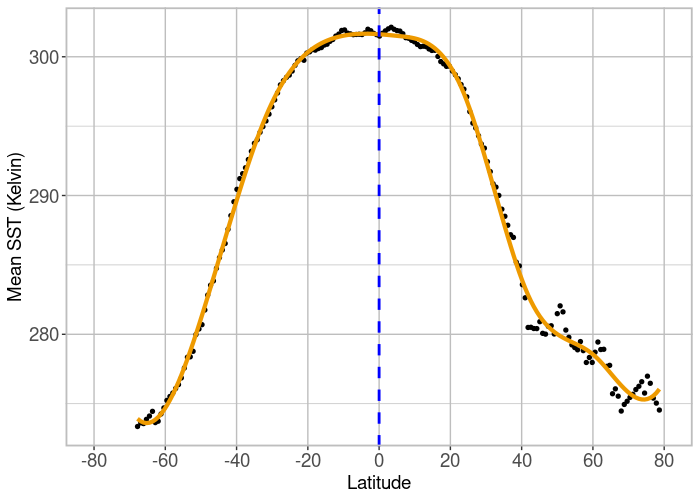}
    \caption{
    Averaged binned SST in 180 latitudinal bins (black dots) with the cubic smoothing spline fit superimposed (orange line).}
    \label{fig:basis_funs_11}
\end{figure}

\begin{figure}[!htb]
    \centering
    \includegraphics[width=\textwidth]{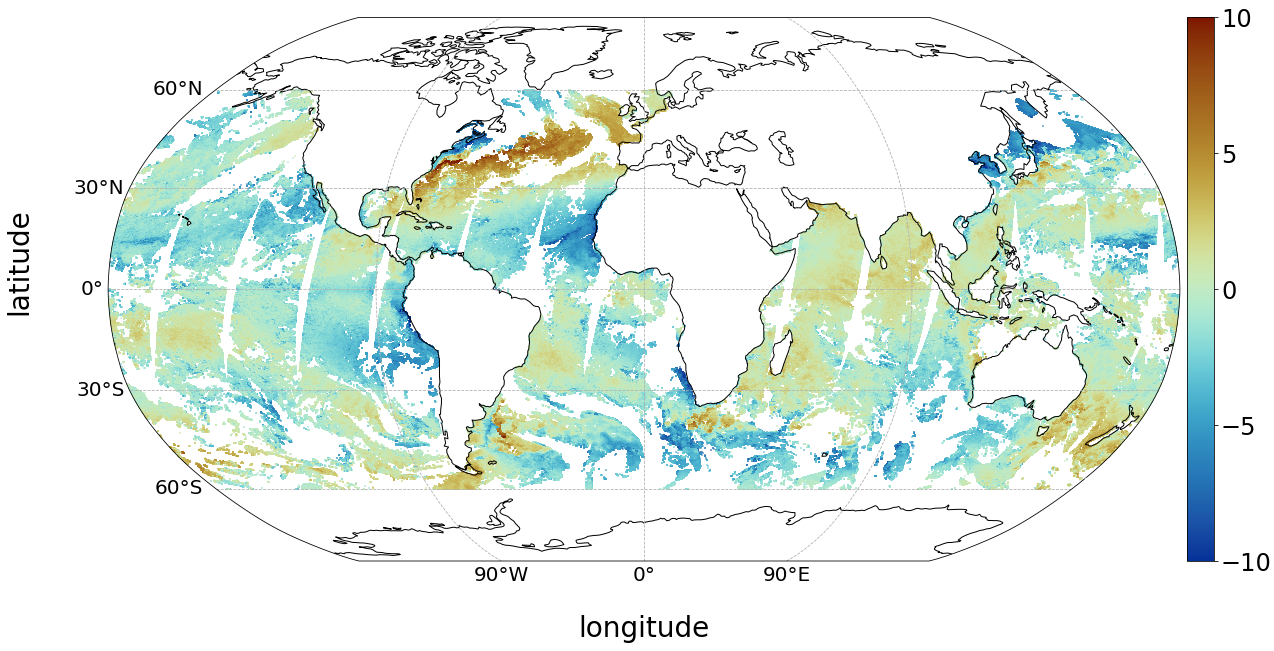}
    \caption{SST residuals on April 7, 2019. The removed mean is estimated by cubic smoothing splines using 11 basis functions.}
    \label{fig:res_field}
\end{figure}

    We denote the SST at location $\bs\in\domain$ by $z(\bs)$. We model $z(\bs) = \mu(\bs)+y(\bs)$, where $\mu(\bs)$ is the mean structure to be estimated so that the resulting residual random field $y(\bs)$ can be assumed a zero-mean Gaussian process. 
We tested a large suite of trend models, including (local) polynomial regressions, thin-plate splines, and cubic smoothing splines. Of these models, the cubic smoothing spline class of models as a function of latitude yields the best balance between accurate estimation of the mean and over-fitting. Ten-fold cross-validation for models with the number of cubic smoothing spline basis functions ranging from five to fifteen is used to select the optimal model. In each fold in the cross-validation, we bin the training data into 180 latitudinal bins and compute the mean SST of each bin. We fit with a number of cubic smoothing splines to these binned data and predict the SST on the validation locations. Figure \ref{fig:cv_spline_fit} in the Supplementary Material shows the results of the cross-validation error versus the selected number of basis functions.
We select 11 cubic smoothing spline functions for their relatively low degrees of freedom and the ability to accurately approximate the mean function $\mu(\bs)$.
Figure \ref{fig:basis_funs_11} depicts the fit using the 11 selected basis functions.
Because the observations are very sparse for latitudes greater than $60^\circ$, we choose our study area as $[180^\circ\textrm{W}, 180^\circ\textrm{E}] \times [60^\circ\textrm{S}, 60^\circ\textrm{N}]$, leading to 43,802,698 observations in total. Figure \ref{fig:res_field} shows the residual spatial field $y(\bs)$ in the study area after the mean trend removal.
More details about the residuals are shown in Figure~\ref{fig:basis_11_more} in the Supplementary Material.

    \subsection{\mra Domain Decomposition\label{subsec:domain_decomp}}
\begin{figure}[!t]
    \renewcommand{\thesubfigure}{\Alph{subfigure}}
    \centering
    \begin{subfigure}{.47\textwidth}
        \subcaption{$m=1$}
        \includegraphics[width=\textwidth,height=3cm]{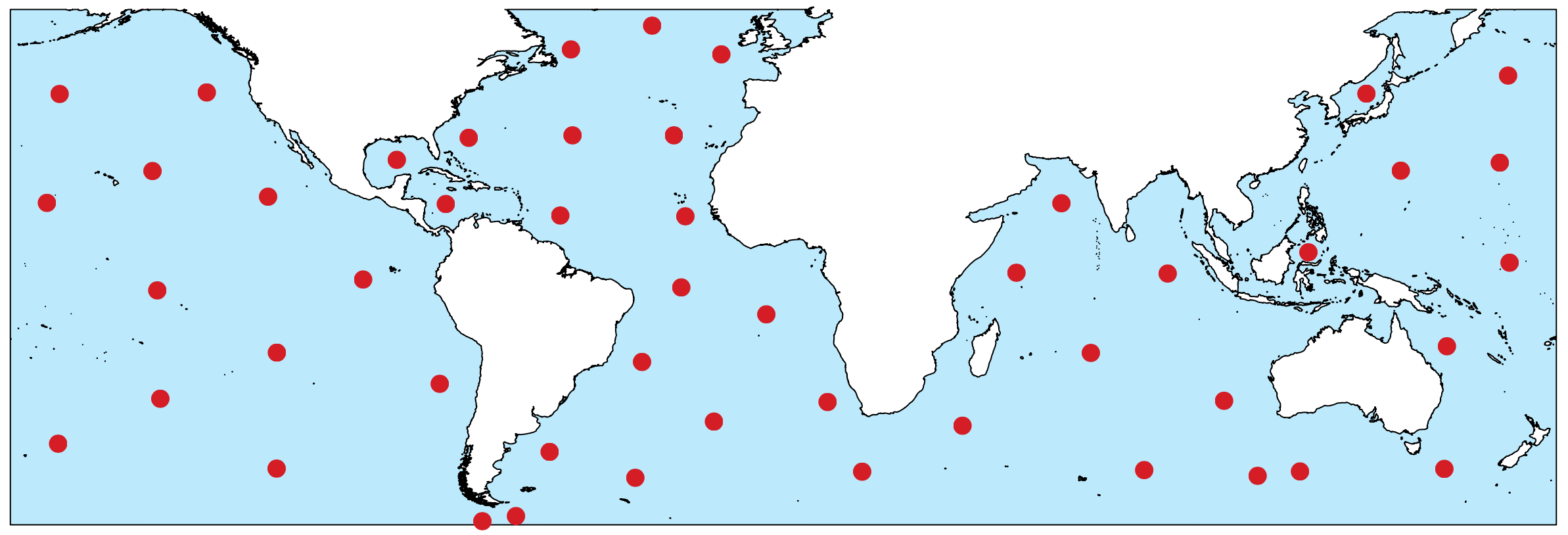}
    \end{subfigure}
    \hspace{5mm}
    \begin{subfigure}{.47\textwidth}
        \subcaption{$m=2$}
        \includegraphics[width=\textwidth,height=3cm]{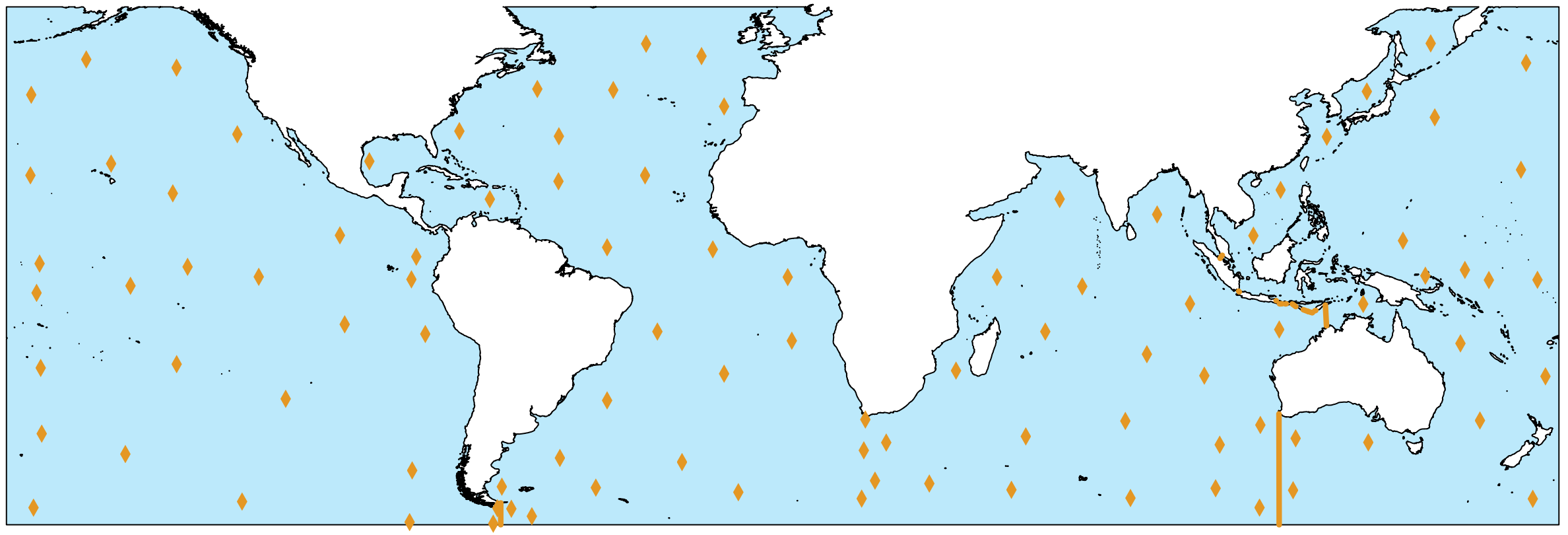}
    \end{subfigure}
    \begin{subfigure}{.47\textwidth}
        \subcaption{$m=3$}
        \includegraphics[width=\textwidth,height=3cm]{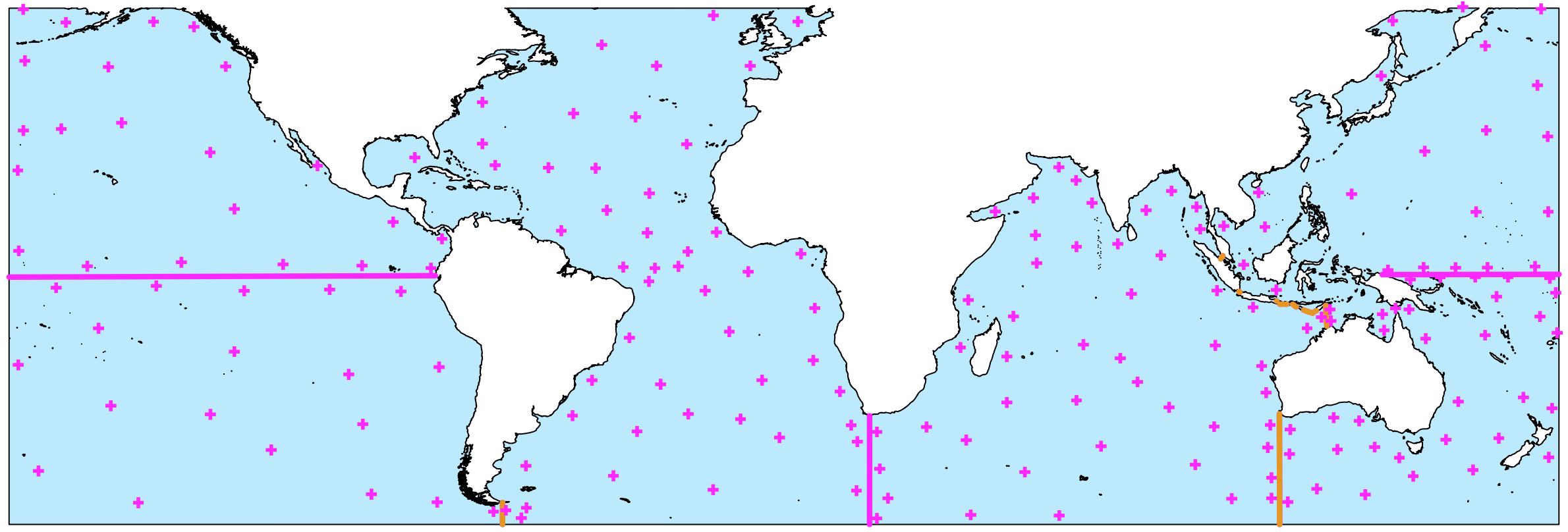}
    \end{subfigure}
    \hspace{5mm}
    \begin{subfigure}{.47\textwidth}
        \subcaption{$m=4$}
        \includegraphics[width=\textwidth,height=3cm]{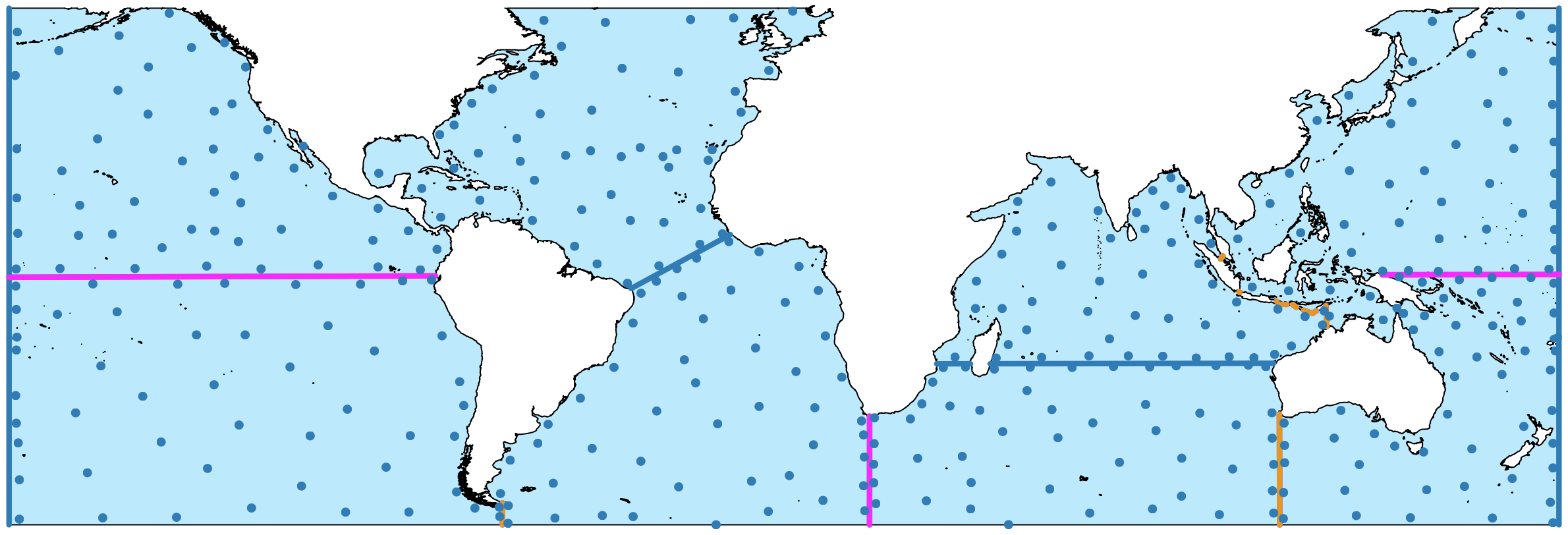}
    \end{subfigure}
    
    \caption{
    First four levels of the hierarchical domain decomposition. Knots and partitions at a given level are shown in the same color.
    In each plot, we show the knots at the current level and the boundaries of the current and coarser partitions.
    Partitions are not placed on land as we only consider SST on open water and therefore land masses are not part of the domain.
    \label{fig:GisDomainDecomp}}
 \end{figure}

    To perform \mra inference for our global SST data, we hierarchically decompose the spatial domain using GIS software~\citep{desktop2011release} in a manner such that we consider realistic spatial dependence informed by physical boundaries and oceanographic knowledge.
Broadly speaking, we place knots and partitions in a way that weakens dependence across land barriers, by placing partition boundaries along (i.e., on top of) land, and avoiding knot placement at lower resolutions too close to land barriers.
We choose $M=15$ levels in total.
For level $m = 2, \ldots, M-1$, each region in level $m-1$ is split into $J=2$ subregions as constructed regions in level $m$.
We place $r=49$ knots in each region for level $m < M$.
We use global landmasses accurate to 10 km resolution based on the World Geodetic System 1984 (WGS84) coordinate system to create the ocean map for the study area between latitudes $60^\circ$S and $60^\circ$N.
We only place the knots on the open ocean, which, as customary for SST products, excludes smaller bodies of water (i.e., Hudson Bay, the Mediterranean Sea, the Baltic Sea, the Black Sea, the Caspian Sea, the Red Sea, and the Persian Gulf).
Placement of knots for levels $m= 1, \ldots, 10$ and region boundaries for $m= 2, \ldots, 10$ are manually performed in the GIS software. In the ocean,
knots are placed strategically near partition boundaries for both the current and the next level, so that the introduced approximation errors near boundaries are small.
Figure~\ref{fig:GisDomainDecomp} shows the region boundaries and knots for the first four levels. 
For levels $m \geq 11$, the region areas are small (the maximum region area at this level is 
less than $7.34\times 10^5$ km$^2$), and most of the large-scale spatial correlation structure has been adequately captured in coarser levels.
Therefore, for simplicity, we split regions perpendicular to the longer spatial dimension at the coordinate mean and place knots randomly in each subregion.
To capture the very fine-scale variation, the observation locations are used as knots in the finest level $m = M = 15$. 
To avoid slow computation for regions that happen to contain a large number of observations, we do not fix $J=2$ but keep partitioning the regions in $m=14$ until all the obtained subregions, which are regions in $m=15$, contain fewer than 2,000 observations; hence, the number of subregions $J$ varies between regions at $m=14$, with $J=2^k$ for some $k \in \mathbb{N}$ for each region.

    \subsection{Assessing Nonstationarity\label{subsec:assess_nonstat}}
    To specify the nonstationary covariance model, we create a $2^\circ \times 2^\circ$ grid in the study area 
$ [180^\circ\textrm{W}, 180^\circ\textrm{E}] \times [60^\circ\textrm{S}, 60^\circ\textrm{N}]$ and obtain the local estimates of $\bftheta(\bs^\ast)$ via the procedure described in Section~\ref{subsec:nonstat_modeling}.
Grid points on lands are removed.
We denote by $\texttt{lon}(\bs^\ast)$ and $\texttt{lat}(\bs^\ast)$ the associated longitude and latitude for $\bs^\ast$.
At each grid point $\bs^\ast$, $B_1(\bs^\ast)$ and $B_2(\bs^\ast)$ are chosen as $B_1(\bs^\ast)=[\texttt{lon}(\bs^\ast) - 2^\circ, \texttt{lon}(\bs^\ast) + 2^\circ] \times [\texttt{lat}(\bs^\ast) - 2^\circ, 
\texttt{lat}(\bs^\ast) + 2^\circ]$ and $B_2(\bs^\ast)=[\texttt{lon}(\bs^\ast) - 20^\circ, \texttt{lon}(\bs^\ast) + 20^\circ] \times [\texttt{lat}(\bs^\ast) - 20^\circ, 
\texttt{lat}(\bs^\ast) + 20^\circ]$. We choose $N_s = 800$ and $N_b = 100$ for the observations in $B_1(\bs^\ast)$ and $B_2(\bs^\ast)\backslash B_1(\bs^\ast)$, respectively.
We observe noisy estimates if the number of chosen observations in a small box $B_1(\bs^\ast)$ is insufficient due to a gap in the satellite data. Therefore, we eliminate all the estimates at lattice points $\bs^\ast$ where the number of available observations in $B_1(\bs^\ast)$ is smaller than 800.
Figure~\ref{fig:smoothness} in the Supplementary Material shows the local estimates of the smoothness parameter from April 1 to 7, 2019.
We observe that the variability of the smoothness estimates is small, which meets the requirement that a fixed smoothness parameter is used in the nonstationary model~\eqref{eq:paciorek_and_schervisch}.
As the majority of the estimates are around 0.5, we fix $\nu = 0.5$, and thus $R^{S}(\cdot)$ in Equation~\eqref{eq:paciorek_and_schervisch} is the exponential function, which brings computational benefits by avoiding the expensive evaluation of the Bessel function.
The nonstationary covariance function now becomes (see the derivation in the Appendix): 
\begin{equation}
\label{eq:nonstat_exp}
C(\bs,\bs') = 
\sigma(\bs)\sigma(\bs')
\left(\dfrac{2\beta(\bs)\beta(\bs')}
{\beta^2(\bs)+\beta^2(\bs')}\right)^{3/2}
\exp
\left(-
  \dfrac
    {\|\bs-\bs'\|}
    {\sqrt{
        \frac{1}{2}
        \big(
            \beta^2(\bs)+\beta^2(\bs')
        \big)
        }
    }
\right) + \tau^2(\bs)\mathbbm{1}_{\{\bs=\bs'\}}.
\end{equation}

We then locally estimate the parameters $\sigma(\cdot)$, $\beta(\cdot),$ and $\tau(\cdot)$ at lattice points again with the exponential covariance function. Figure~\ref{fig:local_estimate} in the Supplementary Material depicts all the local estimates $\hat \bftheta(\bs^\ast)$ from April 1 to 7, 2019 for the nonstationary covariance function~\eqref{eq:nonstat_exp}.
We observe that the patterns during the week are consistent, suggesting that the spatial covariance structure remains similar during the short time. 

\begin{figure}[!htb]
    \centering
    \includegraphics[width=\textwidth]{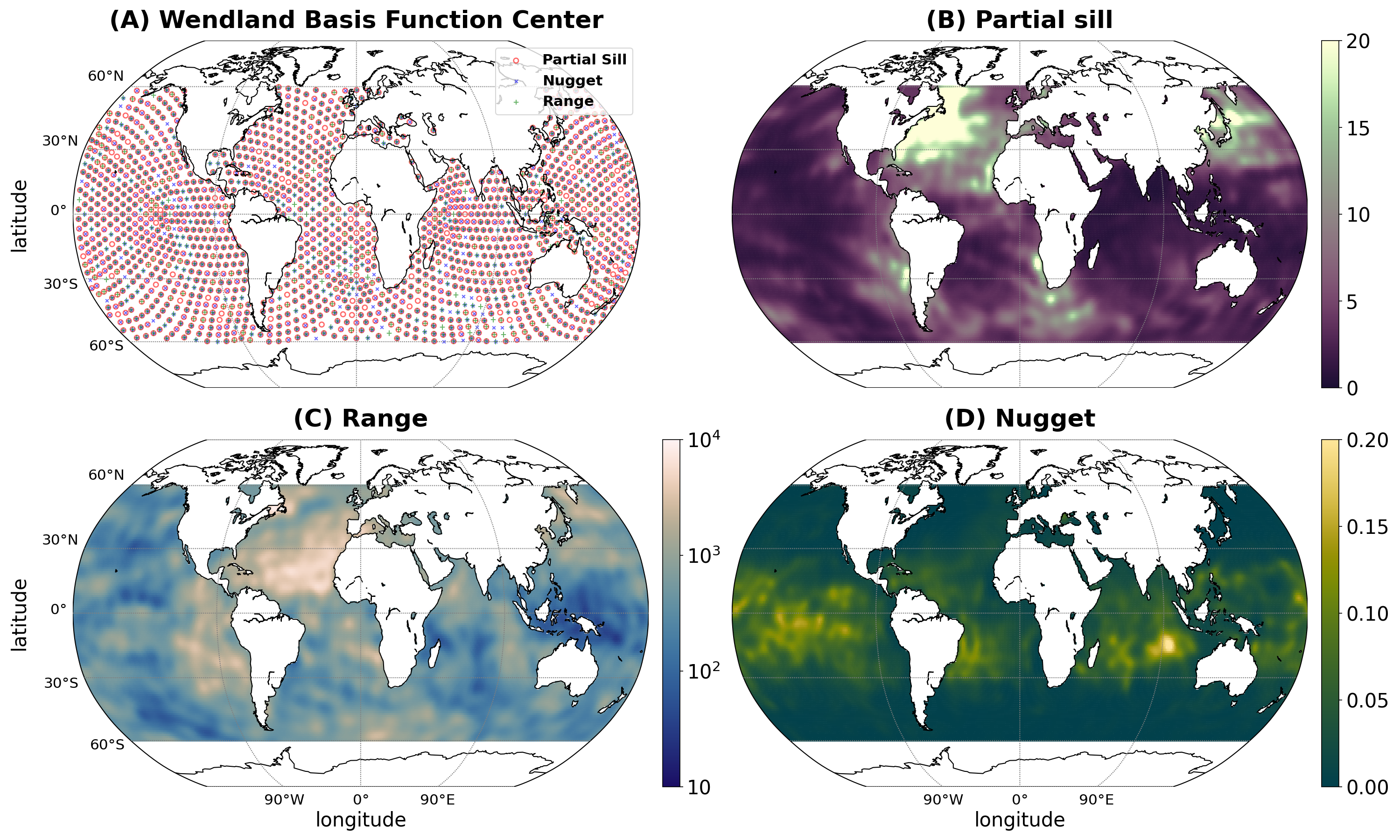}
    \caption{(A): The Wendland basis function center locations with non-zero coefficients. (B)--(D): Spatially varying sill $\sigma^2(\bs)$, range $\beta(\bs)$ in km, and nugget $\tau^2(\bs)$ estimates in the exponential covariance function built from 1399, 1239, 1337 Wendland basis functions, respectively.}
    \label{fig:sp_paras}
\end{figure}

As explained in Section~\ref{subsec:nonstat_modeling}, 
we smooth the local parameter estimates by regressing the estimates onto a set of Wendland basis functions.
To compensate for data vacancy in some areas due to gaps on April 7, we stack all the parameter estimates from April 1 to 7.
With exploratory analysis, we find that logarithmic transformations to the parameter estimates (which must be positive) make them closer to normality, so we perform the regression on the logarithmic scale.
We place Wendland basis functions centered at $\tilde\bs$ in the study area $[-180^\circ\textrm{W}, -180^\circ\textrm{E}] \times [60^\circ\textrm{S}, 60^\circ\textrm{N}]$ and remove $\tilde\bs$ falling on land, so that basis functions only obtain their maximum on the ocean.
By cross-validation of $N_w$ and $\ell$, we obtain 1399, 1239, and 1337 non-zero basis-function coefficients for the partial sill, range, and nugget, respectively, that lead to optimal smoothing results. The center locations of the selected basis functions are shown in Figure~\ref{fig:sp_paras}(A).
Figure~\ref{fig:sp_paras} also illustrates the spatially varying parameter estimate map built from the Wendland basis functions.
These spatially varying parameters are used in the covariance function \eqref{eq:nonstat_exp} to be approximated by the \mra.
    
    \subsection{Prediction Performance Assessment\label{subsec:result}}
\begin{figure}[!t]
    \centering
    \includegraphics[width=8.75cm,height=5cm]{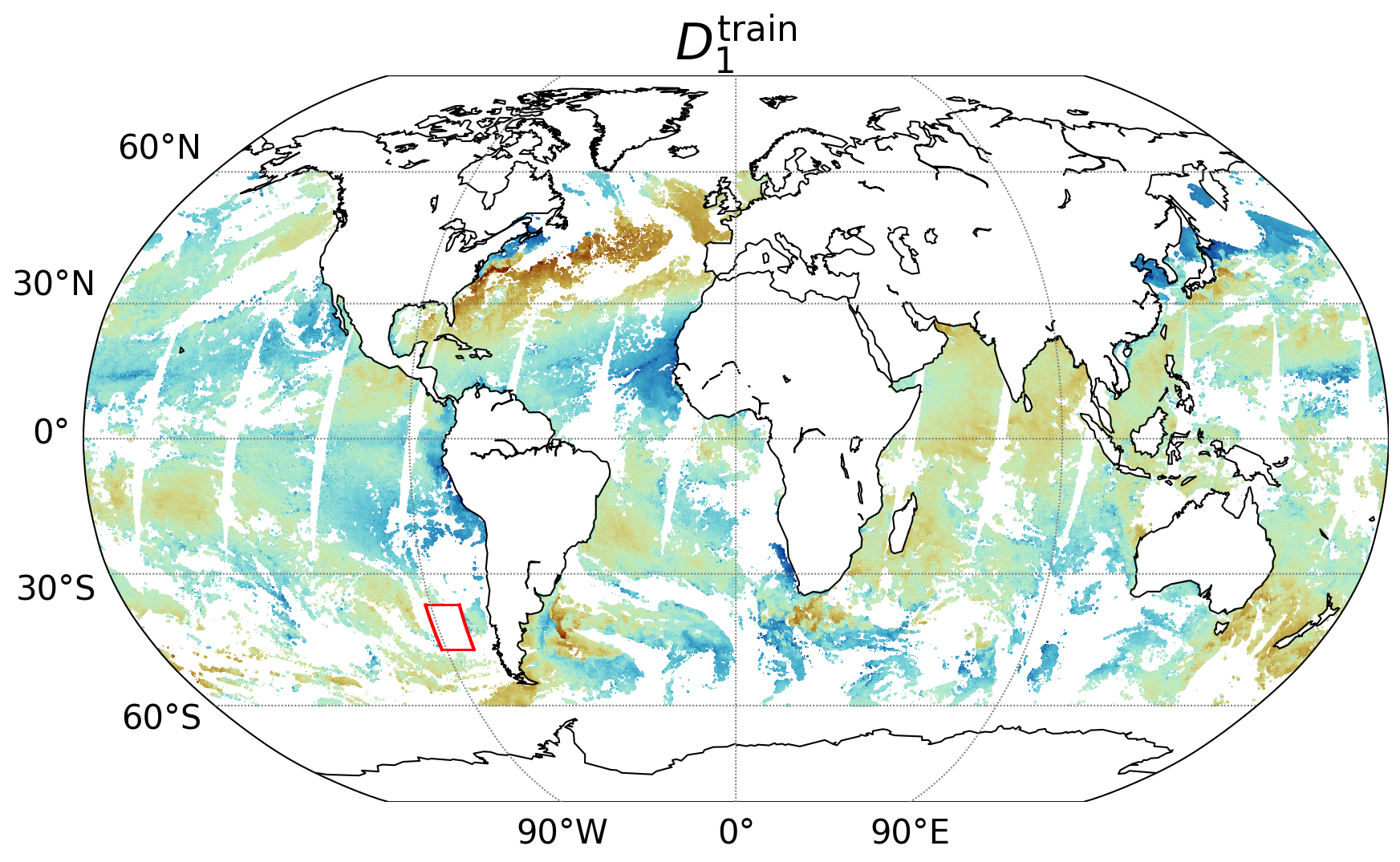}
    \includegraphics[width=5cm,height=5cm]{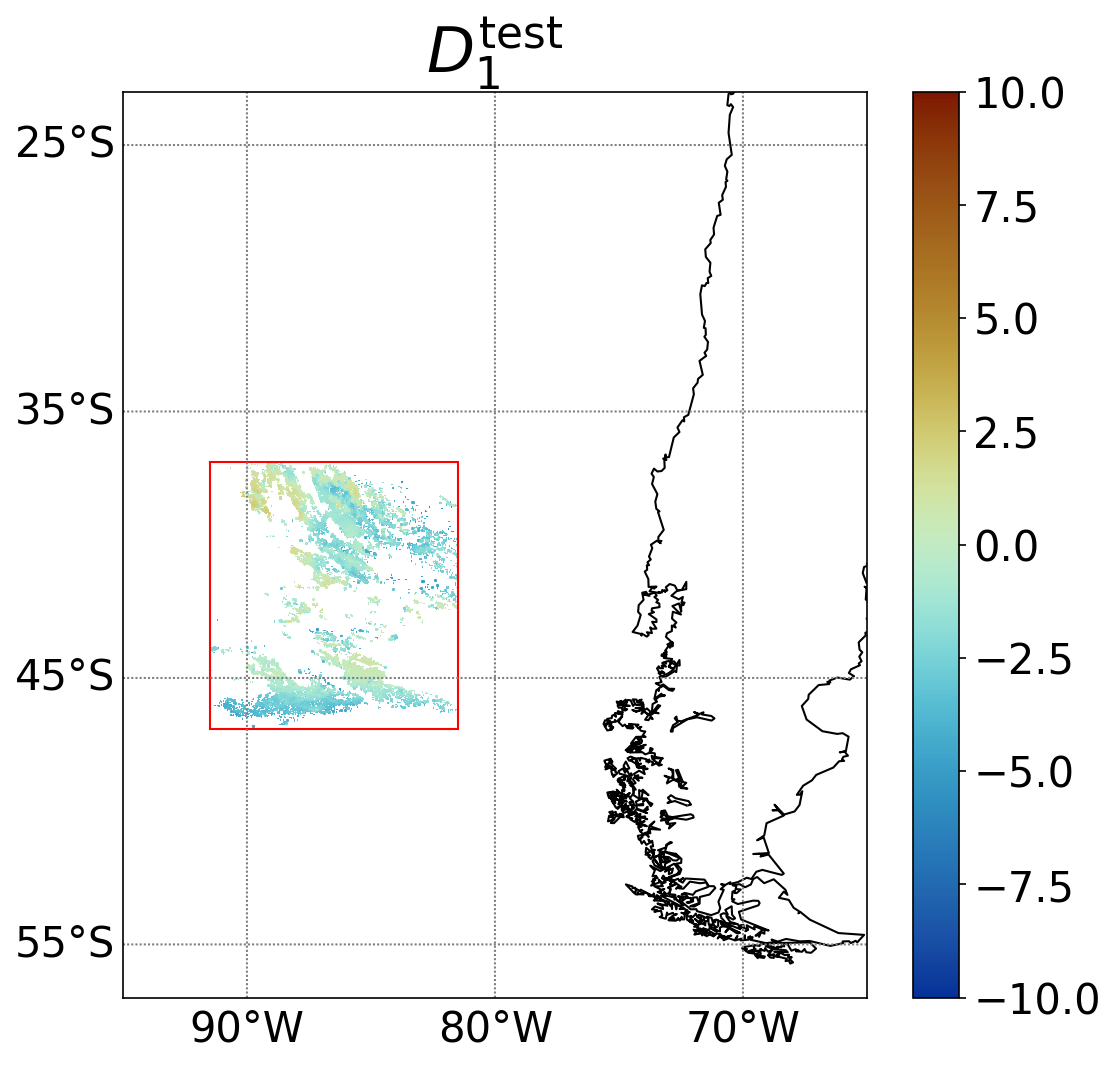}
    \caption{Training and testing data (red box) for Gap 1, $D^\textrm{train}_1, D^\textrm{test}_1$.}
    \label{fig:gap1_data}
\end{figure}

    A common feature in satellite datasets is large gaps that exist due to the inability to obtain measurements under certain atmospheric conditions such as cloud cover. To provide a complete high-resolution data product, the ability to fill large gaps with accurate (probabilistic) predictions is of great importance. 
Therefore, we randomly place large gaps and hold out the observations within the gaps to assess prediction performance.
More precisely, we randomly sample a location (lon, lat) in the ocean and build a gap centered at the location with size $10^\circ\times10^\circ$ as $B = [\textrm{lon} - 5^\circ, \textrm{lon} + 5^\circ]\times
[\textrm{lat} - 5^\circ, \textrm{lat} + 5^\circ]$.
We hold out all the observations in $B$ as testing data $D^\textrm{test}$ and use the rest as training data  $D^\textrm{train}$, i.e., $D^\textrm{test} = \{y(\bs_i): \big(\texttt{lon}(\bs_i), \texttt{lat}(\bs_i)\big)\in B\}$ and $D^\textrm{train} = \{y(\bs_i): \big(\texttt{lon}(\bs_i), \texttt{lat}(\bs_i)\big)\in B^c\}$.
We repeat this procedure 100 times to obtain datasets for 100 independent experiments:
$D^\textrm{train}_1, D^\textrm{test}_1, \ldots, D^\textrm{train}_{100}, D^\textrm{test}_{100}$.
For a more uniform assessment of prediction performance across the 100 experiments, in each experiment $k=1,\ldots,100$, we make sure the number of observations in $B_k$ is greater than 50,000 (by resampling the center point until the criterion is met) and randomly choose 50,000 samples in building the testing dataset.
Figure~\ref{fig:gap1_data} shows one example, the training and testing dataset for Gap 1, $D^\textrm{train}_1, D^\textrm{test}_1$.

In each experiment $k$, we perform the \mra with domain configurations shown in Section~\ref{subsec:domain_decomp} to obtain predictive distributions of all locations in $D^\textrm{test}_k$ conditional on $D^\textrm{train}_k$ with the mean structure described in Section~\ref{subsec:mean} and covariance model~\eqref{eq:nonstat_exp}.
We use the MSPE to assess the point prediction performance and the log-score and Continuous Ranked Probability Score (CRPS) to assess the probabilistic prediction performance.
The log-score and CRPS are strictly proper scoring rules for the assessment of probabilistic prediction quality~\citep[the scores attain the minimum when the realized observations are exactly from the predictive distribution;][]{GR2007}. 
For Gaussian predictive distributions with prediction mean $\hat y(\bs)$ and standard deviation $\hat \varsigma(\bs)$ at location $\bs$, the log-score and CRPS have the form as follows~\citep[e.g.,][]{GK2014},
\[
\begin{array}{rcl}
    \textrm{Log-score}(\bs)  & = &
    \big[\log(2\pi)+{\{y(\bs)-\hat y(\bs)\}^2}/{\hat\varsigma^2(\bs)}\big]/2
+\log\big(\hat\varsigma(\bs)\big), \\
    \textrm{CRPS}(\bs)  & = &
    \{y(\bs)-\hat y(\bs)\}
    \Big[2\Phi\Big(
        \dfrac{y(\bs)-\hat y(\bs)}{\hat\varsigma(\bs)}
        \Big)
    -1\Big]
+2\hat\varsigma\phi\Big(
    \dfrac{y(\bs)-\hat y(\bs)}{\hat\varsigma(\bs)}
\Big)
-\dfrac{\hat\varsigma(\bs)}{\sqrt{\pi}},
\end{array}
\]
where $\Phi(\cdot)$ and $\phi(\cdot)$ are the standard normal cumulative distribution function and probability density function, respectively. 
In each experiment, the log-score and CRPS reported are the mean of the corresponding metrics over all the testing locations.

\begin{figure}[!t]
    \centering
\includegraphics[width=0.95\textwidth]{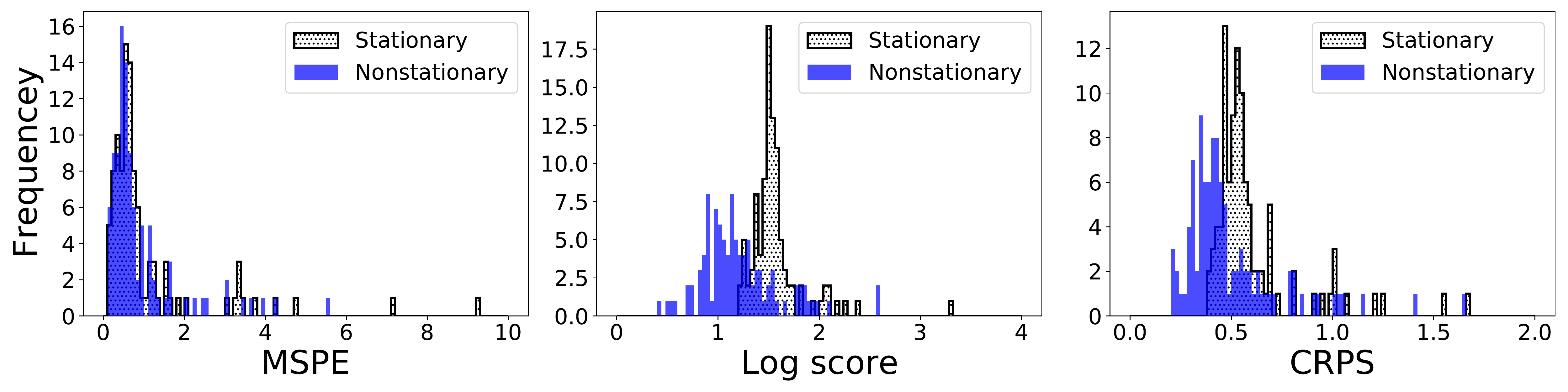}
    \caption{Histogram of MSPE, log-score, and CPRS from the stationary (filled with black dots) and nonstationary (solid blue) models in the 100 experiments.}
    \label{fig:summary}
\end{figure}

\begin{figure}[!t]
    \centering
    \includegraphics[width=.4\textwidth]{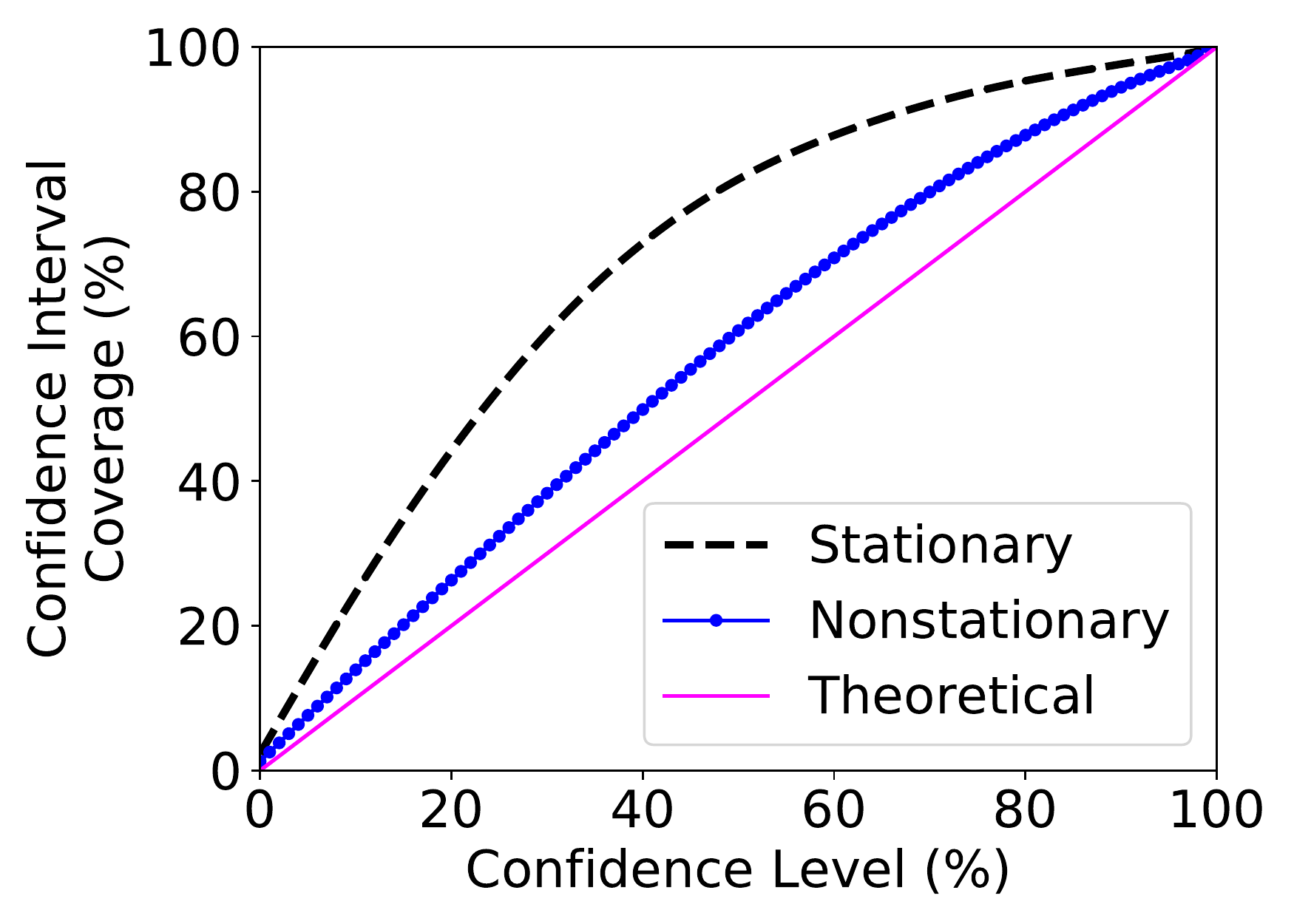}
    \caption{Confidence interval coverage from the stationary (black dashed line) and nonstationary (dotted blue line) models in the 100 experiments. The solid magenta line shows the theoretical coverage from the nominal level.}
    \label{fig:CI}
\end{figure}

For comparison, we also show the results from the stationary \mra model.
For the stationary model, we use the \mra first to estimate the four stationary covariance model parameters via maximum likelihood estimation and then obtain predictions based on the estimated stationary \mra model.
The MSPE, log-score, and CPRS from the stationary and nonstationary models in the 100 experiments are summarized in Figure~\ref{fig:summary} and Table~\ref{tab:summary}.
We observe that incorporation of nonstationarity yields more accurate point prediction and substantially improves the predictive distribution.
We also provide an intuitive way to illustrate the improvement of the probabilistic forecast in Figure~\ref{fig:CI}, where we calculate the proportion of testing observations falling into the interval obtained by the Gaussian predictive distribution with the nominal level. 
Though slightly inflating the variance, the nonstationary model has confidence interval coverage substantially closer to the theoretical values than the stationary model for all levels. 

\begin{table}[!t]
    \centering
    \caption{Average of MSPE, log-score, and CPRS of the corresponding metrics over the 100 experiments for the stationary and nonstationary models, respectively.
    ``\#gaps nonstationary better'' provides the number of experiments where the nonstationary model improves upon the stationary model with respect to the corresponding metric.
    }
    \begin{tabular}{|c|c|c|c|}
    \hline
        Model & MSPE & Log score & CRPS \\ \hline
        
        
        
        Stationary & 1.08 & 1.57 & 0.60 \\
        Nonstationary & 1.01 & 1.21 & 0.49\\
        \hline
        \#gaps nonstationary better & 62 & 96 & 92 \\
        \hline
    \end{tabular}
    \label{tab:summary}
\end{table}

    \subsection{High-Resolution Product}
\begin{figure}[!t]
    \centering
    \includegraphics[width=1\textwidth]{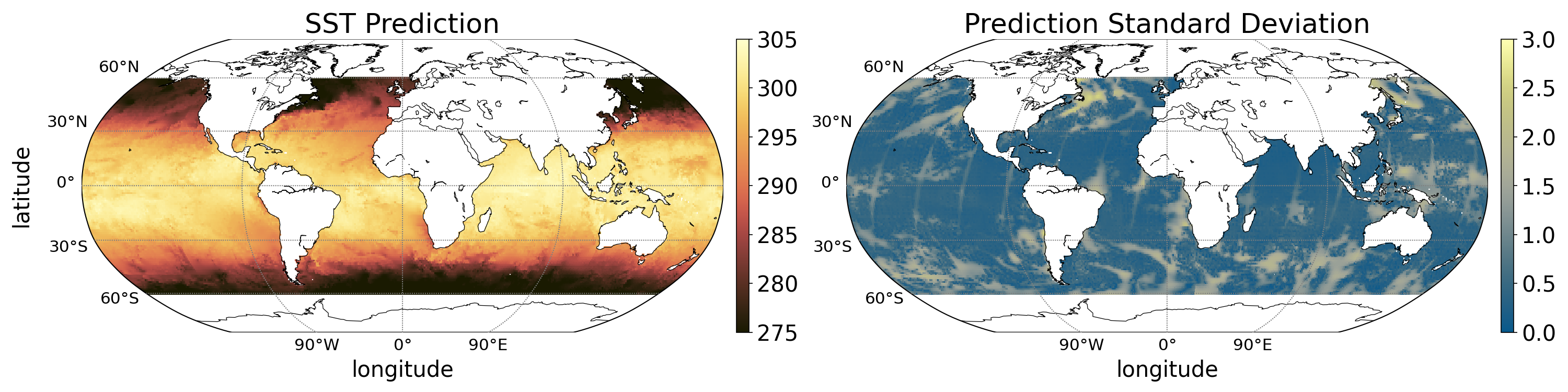}
    \caption{High-resolution SST product using the nonstationary \mra model.}
    \label{fig:product}
\end{figure}

    The high-resolution SST product for the entire study area using the nonstationary \mra model is shown in Figure~\ref{fig:product}.
The complete map of prediction means  resolves SST variation even on very fine scales and does not show obvious artifacts on boundaries between partitions, implying a good domain decomposition scheme in our analysis. The prediction standard deviations reflect the missing data pattern and show nonstationary structures across the globe (e.g., large values in the ocean near the south of the Labrador Sea).
The \mra prediction is performed on 25 nodes on the Cheyenne supercomputer~\citep{Cheyenne}, each of which has 36
2.3-GHz Intel Xeon cores, and takes 35 seconds (each node launches one MPI process and 36 OpenMP threads).

\section{Summary and Future work\label{sec:discuss}}
Addressing the urgent need to provide statistically rigorous full-coverage global data products based on exceedingly massive satellite observations, we have developed a nonstationary version and distributed-memory implementation of the \mra modeling framework. We employ this method, following a careful analysis of the mean and nonstationary covariance structure, to provide a daily global SST product based on 43 million observations. The prediction performance of the nonstationary model is substantially better compared to the stationary model and highlights the need to incorporate nonstationarity when dealing with global geophysical data.
Additional future satellite instruments with increasingly high resolution are expected to strengthen the need for the kind of methodology provided here.

One particularly attractive feature of a statistically rigorous approach is the possibility of extending the SST product into the form of joint probability distributions, which in turn enables us to resolve SST gradients with proper uncertainties at any location in the ocean. 
This is important for the study of atmosphere-ocean coupling, and it allows for answering scientific questions regarding, for example, the effect of climate change on upwelling and the implications on coastal fisheries. 
We plan on conducting such studies in collaboration with oceanographers in future research.

\if1\blind
{
\section*{Acknowledgments}

MK was partially supported by National Science Foundation (NSF) Grants DMS--1521676, DMS--1654083, and DMS--1953005, and by the National Aeronautics and Space Administration (80NM0018F0527). We would like to thank Samuel H. Blake for assistance with GIS. We would like to acknowledge high-performance computing support from Cheyenne (doi:10.5065/D6RX99HX) provided by NCAR's Computational and Information Systems Laboratory, sponsored by the National Science Foundation.
}
\fi


\section*{Appendix \label{app:proofs}}
{
\footnotesize
We derive the form for the nonstationary exponential covariance function given in Equation~\eqref{eq:nscov}. 
We let $\bfSigma_\bs = \beta^2(\bs)\bI_3, \bfSigma_{\bs^\prime} = \beta^2(\bs^\prime)\bI_3$ and plug into Equation~\eqref{eq:paciorek_and_schervisch} and we have:
\[
R^{NS}(\bs, \bs^\prime) = 
    \left(\beta(\bs)^6\right)^{\frac{1}{4}}
    \left(\beta(\bs^\prime)^6\right)^{\frac{1}{4}}
    \left(\frac{2}{\beta^2(\bs) + \beta^2(\bs^\prime)}\right)^{{3}/{2}}
    R^{S}\left( \bs-\bs'; \dfrac{\beta^2(\bs)+\beta^2(\bs')}{2}\bI_3\right).
\]
Considering $R^{S}(\bh; \bfSigma) = \big(\sqrt{2\nu\bh^\top\bfSigma^{-1}\bh}\big)^\nu K_\nu\big(\sqrt{2\nu\bh^\top\bfSigma^{-1}\bh}\big) / \{\Gamma(\nu)2^{\nu-1}\}$, we have
\[
(\bs-\bs')^\top\left(\dfrac{\beta^2(\bs)+\beta^2(\bs')}{2}\bI_3\right)^{-1}(\bs-\bs')
=
\dfrac{2\|\bs-\bs'\|^2}
{\beta^2(\bs)+\beta^2(\bs')}
\]
and subsequently
\[
    R^{NS}(\bs, \bs^\prime) =  
    \left(\dfrac{2\beta(\bs)\beta(\bs')}
{\beta^2(\bs)+\beta^2(\bs')}\right)^{3/2}
\left(
  \dfrac
    {2\sqrt{\nu}\|\bs-\bs'\|}
    {\sqrt{
        \beta^2(\bs)+\beta^2(\bs')
        }
    }
\right)^\nu
K_\nu
\left(
  \dfrac
    {2\sqrt{\nu}\|\bs-\bs'\|}
    {\sqrt{
            \beta^2(\bs)+\beta^2(\bs')
        }
    }
\right).
\]
The mapping $(\bs, \bs^\prime) \mapsto \sigma(\bs)\sigma(\bs^\prime)$ is positive definite \citep[e.g.,][]{berg1984harmonic}. Products of positive definite kernels are positive definite. After adding an independent white noise with variance $\tau^2(\bs)$, we get the valid nonstationary exponential covariance function given in Equation~\eqref{eq:nscov}.
}
\bibliographystyle{apalike}
\bibliography{mendeley,additionalrefs}

\renewcommand{\thesection}{S\arabic{section}}
\renewcommand{\thefigure}{S\arabic{figure}}
\renewcommand{\thetable}{S\arabic{table}}

\clearpage
\begin{center}
	\bf \Huge Supplementary Material
\end{center}

\setcounter{figure}{0}
\setcounter{table}{0}
\setcounter{section}{0}
\setcounter{page}{1}

\begin{figure}[!htb]
    \centering
    \includegraphics[width=\textwidth]{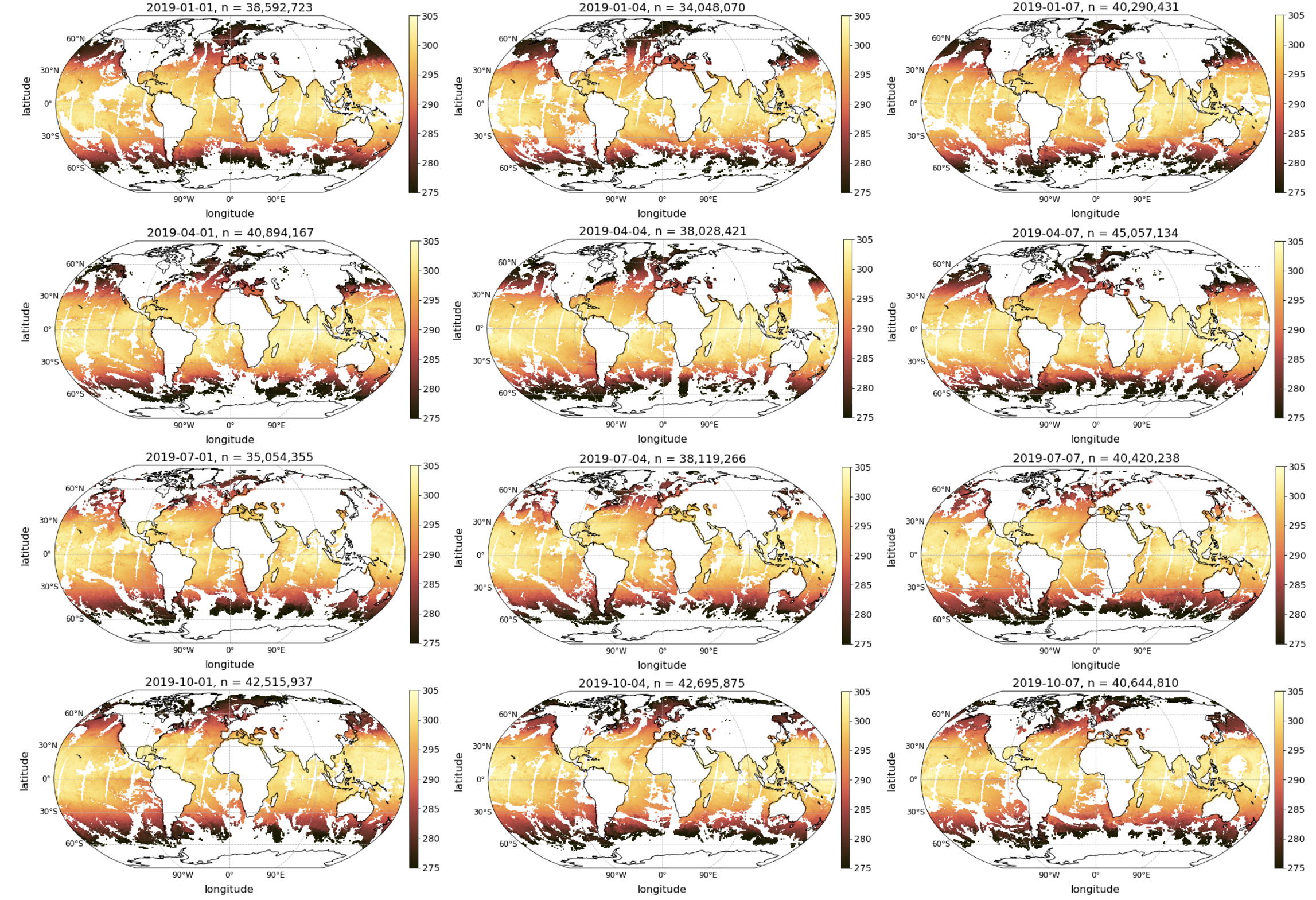}
    \caption{Global nighttime SST for the first, fourth, and seventh days of January, April, July and October 2019. Dates are presented in YYYY-MM-DD format. Rows correspond to months and columns to days. The number of all observations available for each day  ($n$) is presented in the titles. Scale ranges from 275 to 305 Kelvin.}
    \label{fig:globalL2Daily2019_low_res}
\end{figure}

\begin{figure}[!htb]
	\centering
	\includegraphics[width=0.66\textwidth]{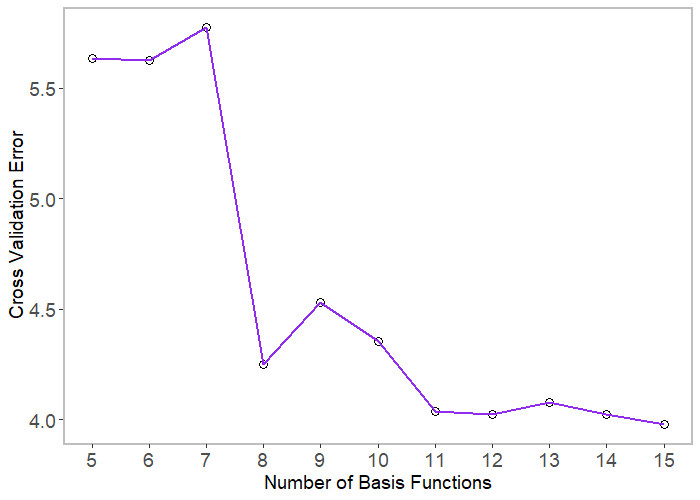}
	\caption{Cross validation error versus number of basis functions (see Section~\ref{subsec:mean}).
	We select 11 cubic smoothing spline functions for their relatively low degrees of freedom and small error.}
	\label{fig:cv_spline_fit}
\end{figure}

\begin{figure}[!htb]
	\centering
	\includegraphics[width=0.95\textwidth]{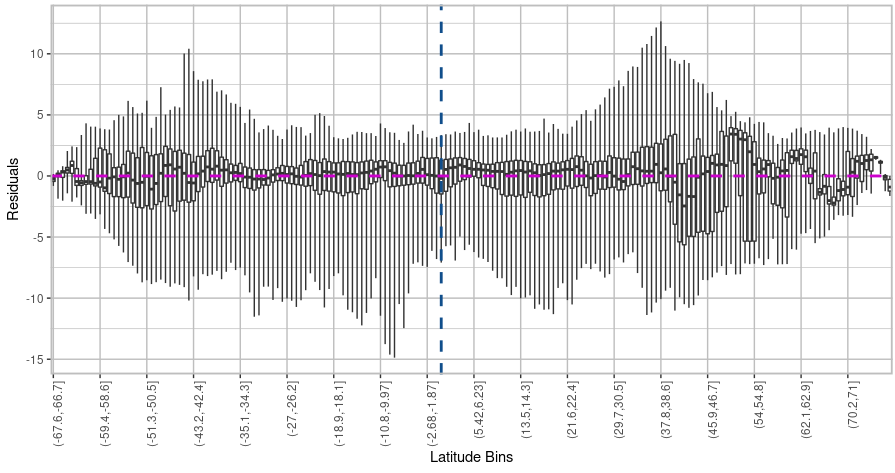}
	\caption{Boxplots of the nighttime SST residuals using 11 cubic smoothing splines aggregated into 180 latitudinal bins (see Section~\ref{subsec:mean}). Bin whiskers are extended to the range of observations. The blue vertical dashed line is placed at the bin containing the equator and the horizontal magenta dashed line is placed at zero. The majority of residuals center at zero, and we observe different variability of residuals at different latitudes, indicating a nonstationary covariance model for the residuals is needed.}
	\label{fig:basis_11_more}
\end{figure}

\begin{figure}[!htb]
    \centering
    \includegraphics[width=\textwidth]{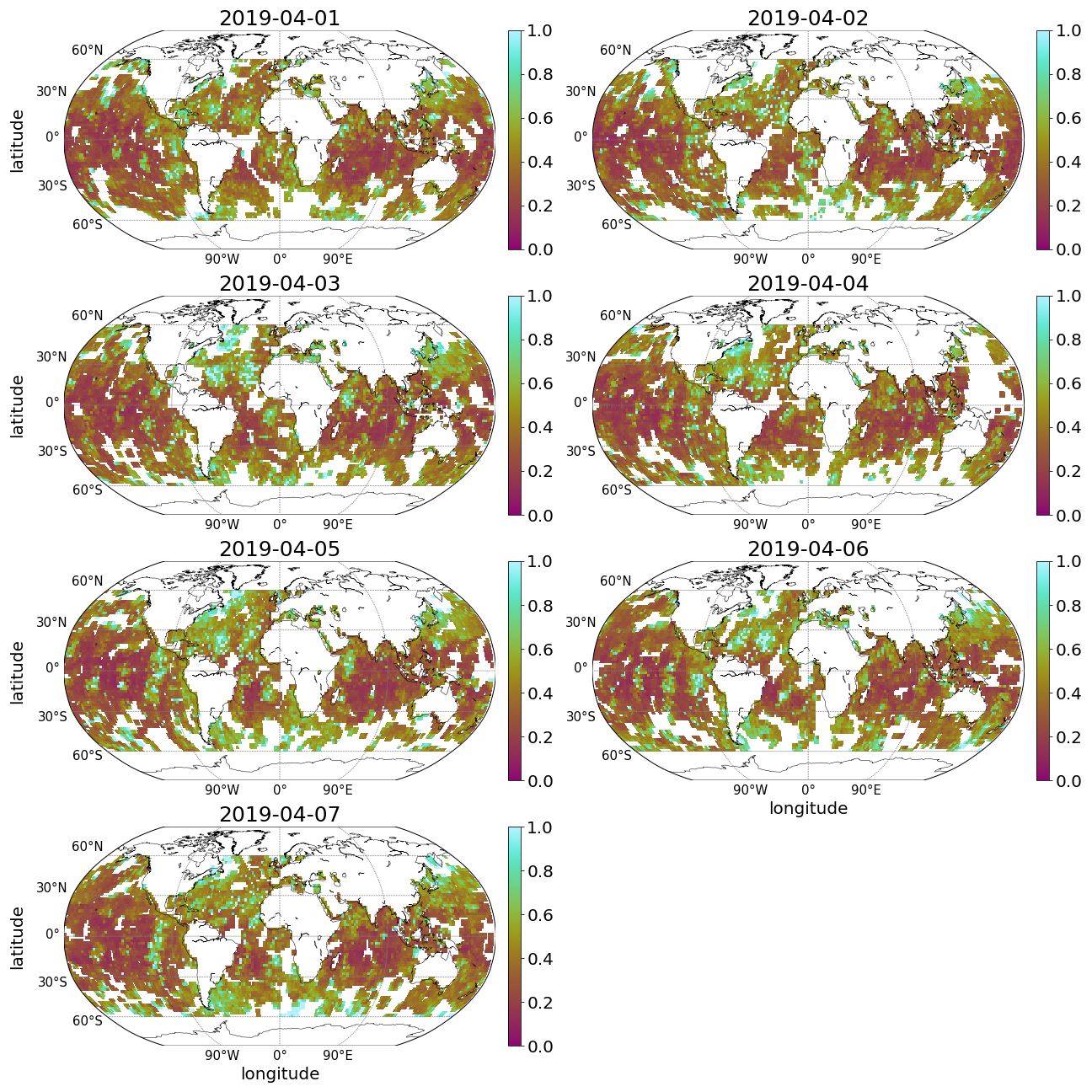}
    \caption{Local estimates of the smoothness parameter from April 1 to 7, 2019.
    Most of them are relatively stable around 0.5, supporting the assumption of exponential covariance being used (see Section~\ref{subsec:assess_nonstat}).}
    \label{fig:smoothness}
\end{figure}

\begin{figure}
	\centering
	\includegraphics[width =.95\linewidth]{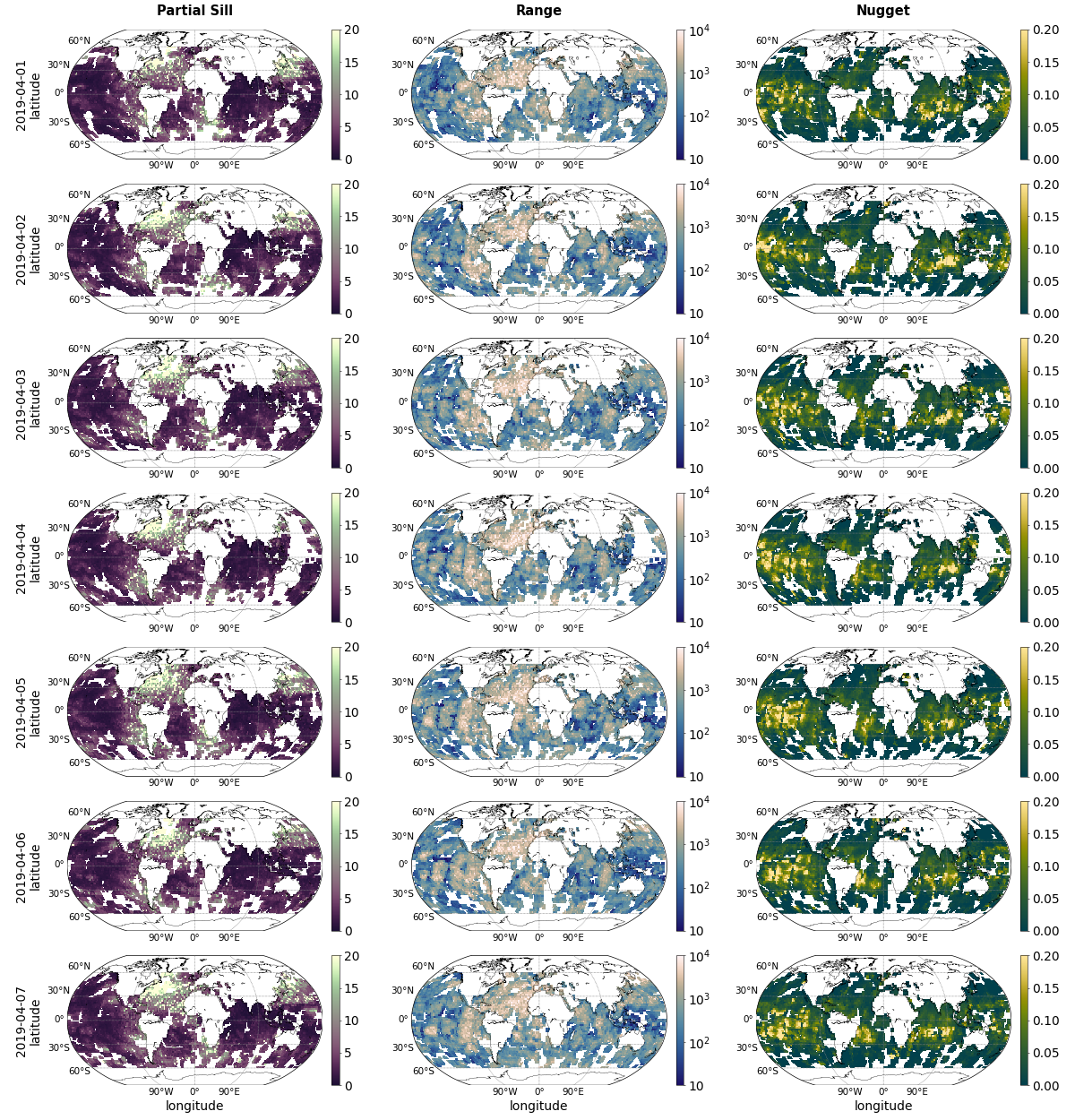}
	\caption{Locally estimated parameters of the exponential covariance function for the nighttime SST residuals from April 1 to 7, 2019.
	Consistent patterns are observed across the week, motivating us to use all of them to obtain smoothed estimates by the Wendland basis functions (see Section~\ref{subsec:assess_nonstat}).}
	\label{fig:local_estimate}
\end{figure}

\end{document}